\input harvmac
\overfullrule=0pt
\abovedisplayskip=12pt plus 3pt minus 3pt
\belowdisplayskip=12pt plus 3pt minus 3pt

\message{S-Tables Macro v1.0, ACS, TAMU (RANHELP@VENUS.TAMU.EDU)}
%
%
\newhelp\stablestylehelp{You must choose a style between 0 and 3.}%
\newhelp\stablelinehelp{You should not use special hrules when stretching
a table.}%
\newhelp\stablesmultiplehelp{You have tried to place an S-Table 
inside another S-Table.  I would recommend not going on.}%
%
%
\newdimen\stablesthinline
\stablesthinline=0.4pt
\newdimen\stablesthickline
\stablesthickline=1pt
%
%
\newif\ifstablesborderthin
\stablesborderthinfalse
\newif\ifstablesinternalthin
\stablesinternalthintrue
\newif\ifstablesomit
\newif\ifstablemode
\newif\ifstablesright
\stablesrightfalse
%
%
\newdimen\stablesbaselineskip
\newdimen\stableslineskip
\newdimen\stableslineskiplimit
%
%
\newcount\stablesmode
\newcount\stableslines
\newcount\stablestemp
\stablestemp=3
\newcount\stablescount
\stablescount=0
\newcount\stableslinet
\stableslinet=0
%
%
%
\newcount\stablestyle
\stablestyle=0
%
%
\def\stablesleft{\quad\hfil}%
\def\stablesright{\hfil\quad}%
%
%
\catcode`\|=\active%
%
%
\newcount\stablestrutsize
\newbox\stablestrutbox
\setbox\stablestrutbox=\hbox{\vrule height10pt depth5pt width0pt}
\def\stablestrut{\relax\ifmmode%
                         \copy\stablestrutbox%
                       \else%
                         \unhcopy\stablestrutbox%
                       \fi}%
%
%
\newdimen\stablesborderwidth
\newdimen\stablesinternalwidth
\newdimen\stablesdummy
\newcount\stablesdummyc
\newif\ifstablesin
\stablesinfalse
%
%
\def\begintable{\stablestart%
  \stablemodetrue%
  \stablesadj%
  \halign%
  \stablesdef}%
\def\stablesadj{%
  \ifcase\stablestyle%
    \hbox to \hsize\bgroup\hss\vbox\bgroup%
  \or%
    \hbox to \hsize\bgroup\vbox\bgroup%
  \or%
    \hbox to \hsize\bgroup\hss\vbox\bgroup%
  \or%
    \hbox\bgroup\vbox\bgroup%
  \else%
    \errhelp=\stablestylehelp%
    \errmessage{Invalid style selected, using default}%
    \hbox to \hsize\bgroup\hss\vbox\bgroup%
  \fi}%
\def\stablesend{\egroup%
  \ifcase\stablestyle%
    \hss\egroup%
  \or%
    \hss\egroup%
  \or%
    \egroup%
  \or%
    \egroup%
  \else%
    \hss\egroup%
  \fi}%
\def\stablestart{%
  \ifstablesin%
    \errhelp=\stablesmultiplehelp%
    \errmessage{An S-Table cannot be placed within an S-Table!}%
  \fi
  \global\stablesintrue%
  \global\advance\stablescount by 1%
  \message{<S-Tables Generating Table \number\stablescount}%
  \begingroup%
  \stablestrutsize=\ht\stablestrutbox%
  \advance\stablestrutsize by \dp\stablestrutbox%
  \ifstablesborderthin%
    \stablesborderwidth=\stablesthinline%
  \else%
    \stablesborderwidth=\stablesthickline%
  \fi%
  \ifstablesinternalthin%
    \stablesinternalwidth=\stablesthinline%
  \else%
    \stablesinternalwidth=\stablesthickline%
  \fi%
  \tabskip=0pt%
  \stablesbaselineskip=\baselineskip%
  \stableslineskip=\lineskip%
  \stableslineskiplimit=\lineskiplimit%
  \offinterlineskip%
  \def\borderrule{\vrule width \stablesborderwidth}%
  \def\internalrule{\vrule width \stablesinternalwidth}%
  \def\thinline{\noalign{\hrule height \stablesthinline}}%
  \def\thickline{\noalign{\hrule height \stablesthickline}}%
  \def\trule{\omit\leaders\hrule height \stablesthinline\hfill}%
  \def\ttrule{\omit\leaders\hrule height \stablesthickline\hfill}%
  \def\tttrule##1{\omit\leaders\hrule height ##1\hfill}%
  \def\stablesel{&\omit\global\stablesmode=0%
    \global\advance\stableslines by 1\borderrule\hfil\cr}%
  \def\el{\stablesel&}%
  \def\elt{\stablesel\thinline&}%
  \def\eltt{\stablesel\thickline&}%
  \def\elttt##1{\stablesel\noalign{\hrule height ##1}&}%
  \def\elspec{&\omit\hfil\borderrule\cr\omit\borderrule&%
              \ifstablemode%
              \else%
                \errhelp=\stablelinehelp%
                \errmessage{Special ruling will not display properly}%
              \fi}%
  \def\stmultispan##1{\mscount=##1 \loop\ifnum\mscount>3 \stspan\repeat}%
  \def\stspan{\span\omit \advance\mscount by -1}%
  \def\multicolumn##1{\omit\multiply\stablestemp by ##1%
     \stmultispan{\stablestemp}%
     \advance\stablesmode by ##1%
     \advance\stablesmode by -1%
     \stablestemp=3}%
  \def\multirow##1{\stablesdummyc=##1\parindent=0pt\setbox0\hbox\bgroup%
    \aftergroup\emultirow\let\temp=}
  \def\emultirow{\setbox1\vbox to\stablesdummyc\stablestrutsize%
    {\hsize\wd0\vfil\box0\vfil}%
    \ht1=\ht\stablestrutbox%
    \dp1=\dp\stablestrutbox%
    \box1}%
  \def\stpar##1{\vtop\bgroup\hsize ##1%
     \baselineskip=\stablesbaselineskip%
     \lineskip=\stableslineskip%
   \lineskiplimit=\stableslineskiplimit\bgroup\aftergroup\estpar\let\temp=}%
  \def\estpar{\vskip 6pt\egroup}%
  \def\stparrow##1##2{\stablesdummy=##2%
     \setbox0=\vtop to ##1\stablestrutsize\bgroup%
     \hsize\stablesdummy%
     \baselineskip=\stablesbaselineskip%
     \lineskip=\stableslineskip%
     \lineskiplimit=\stableslineskiplimit%
     \bgroup\vfil\aftergroup\estparrow%
     \let\temp=}%
  \def\estparrow{\vfil\egroup%
     \ht0=\ht\stablestrutbox%
     \dp0=\dp\stablestrutbox%
     \wd0=\stablesdummy%
     \box0}%
  \def|{\global\advance\stablesmode by 1&&&}%
  \def\|{\global\advance\stablesmode by 1&\omit\vrule width 0pt%
         \hfil&&}%
\def\vt{\global\advance\stablesmode 
by 1&\omit\vrule width \stablesthinline%
          \hfil&&}%
  \def\vtt{\global\advance\stablesmode by 1&\omit\vrule width
\stablesthickline%
          \hfil&&}%
  \def\vttt##1{\global\advance\stablesmode by 1&\omit\vrule width ##1%
          \hfil&&}%
  \def\vtr{\global\advance\stablesmode by 1&\omit\hfil\vrule width%
           \stablesthinline&&}%
  \def\vttr{\global\advance\stablesmode by 1&\omit\hfil\vrule width%
            \stablesthickline&&}%
\def\vtttr##1{\global\advance\stablesmode
 by 1&\omit\hfil\vrule width ##1&&}%
  \stableslines=0%
  \stablesomitfalse}
\def\stablesdef{\bgroup\stablestrut\borderrule##\tabskip=0pt plus 1fil%
  &\stablesleft##\stablesright%
  &##\ifstablesright\hfill\fi\internalrule\ifstablesright\else\hfill\fi%
  \tabskip 0pt&&##\hfil\tabskip=0pt plus 1fil%
  &\stablesleft##\stablesright%
  &##\ifstablesright\hfill\fi\internalrule\ifstablesright\else\hfill\fi%
  \tabskip=0pt\cr%
  \ifstablesborderthin%
    \thinline%
  \else%
    \thickline%
  \fi&%
}%
\def\endtable{\advance\stableslines by 1\advance\stablesmode by 1%
   \message{- Rows: \number\stableslines, Columns:  \number\stablesmode>}%
   \stablesel%
   \ifstablesborderthin%
     \thinline%
   \else%
     \thickline%
   \fi%
   \egroup\stablesend%
\endgroup%
\global\stablesinfalse}
%


\def\tilde{\widetilde}
\def\bar{\overline}
\def\to{\rightarrow}

\def\cN{{\cal N}}

\def\bigone{\hbox{1\kern -.23em {\rm l}}}
\def\ZZ{\hbox{\zfont Z\kern-.4emZ}}
\def\half{{\litfont {1 \over 2}}}

\def\tr{{\rm tr}\,}

\font\litfont=cmr6

\def\cH{{\cal H}}

\font\zfont = cmss10 
\font\litfont = cmr6

\def\bigone{\hbox{1\kern -.23em {\rm l}}}
\def\ZZ{\hbox{\zfont Z\kern-.4emZ}}
\def\half{{\litfont {1 \over 2}}}




\def\frac#1#2{{#1 \over #2}}

\def\cite#1{\#1}

\def\section#1{\newsec {#1}}
\def\subsection#1{\subsec {#1}}
\def\subsubsection#1{\bigskip\noindent{\it #1}}


\def\myI#1 {\int \! #1 \,}

\def\eqn#1#2{\xdef #1{(\secsym\the\meqno)}\writedef{#1\leftbracket#1}%
	$$#2\eqno#1\eqlabeL#1$$%
	\xdef #1{(eq.~\secsym\the\meqno) }\global\advance\meqno by1}

\def\eqr#1{#1 }

\def\spsep{\;\;\;\;\;\;\;\;}   

\def\abs#1{{\vert #1 \vert}}

\def\paren#1{\left( #1 \right)}
\def\brak#1{\left[	#1 \right]}
\def\brac#1{\left\{	#1 \right\}}

\def\ket#1{\vert #1 \rangle}
\def\bra#1{\langle #1 \vert}

\def\comm#1#2{\left[ #1, #2 \right]}

\def\pa{\partial}

\def\cA{{\cal A}}
\def\cF{{\cal F}}

\def\id{{\bf Id}}




\def\ppa#1{\frac \pa {\pa #1}}

\def\PB#1#2{\brac {#1, #2}}

%
\let\useblackboard=\iftrue
%
%
\newfam\black
\def\Title#1#2{\rightline{#1}
\ifx\answ\bigans\nopagenumbers\pageno0\vskip1in%
\baselineskip 15pt plus 1pt minus 1pt
\else
\def\listrefs{\footatend\vskip 1in\immediate\closeout\rfile\writestoppt
\baselineskip=14pt\centerline{{\bf References}}\bigskip{\frenchspacing%
\parindent=20pt\escapechar=` \input
refs.tmp\vfill\eject}\nonfrenchspacing}
\pageno1\vskip.8in\fi \centerline{\titlefont #2}\vskip .5in}

\ifx\answ\bigans\def\tcbreak#1{}\else\def\tcbreak#1{\cr&{#1}}\fi
\useblackboard
\message{If you do not have msbm (blackboard bold) fonts,}
\message{change the option at the top of the tex file.}
\font\blackboard=msbm10 scaled \magstep1
\font\blackboards=msbm7
\font\blackboardss=msbm5
\textfont\black=\blackboard
\scriptfont\black=\blackboards
\scriptscriptfont\black=\blackboardss
\def\Bbb#1{{\fam\black\relax#1}}
\else
\def\Bbb{\bf}
\fi

\def\id{\Bbb {Id}}


\def\cTheta{\Theta}

\def\CSCO{complete set of commuting observables}

\def\suid{\id_{su(2)}}


\def\em {\it}


\Title{\vtop{\hbox{hep-th/0011034}
\hbox{IASSNS-HEP-00/76}}}
{\vbox{\centerline{Non-$\cA$belian Geometry}}}
\centerline{ Keshav Dasgupta\foot{\tt keshav@sns.ias.edu}
 and Zheng Yin\foot{\tt yin@sns.ias.edu}}
\vskip 15pt
\centerline{\it School of Natural Sciences}
\centerline{\it Institute for Advanced Study}
\centerline{\it Einstein Drive}
\centerline{\it Princeton NJ 08540, USA}

\ \smallskip
\centerline{\bf Abstract}

	Spatial noncommutativity is similar and can even be related to 
the non-Abelian nature of multiple D-branes.  
But they have so far seemed independent of each other.
Reflecting this decoupling, the algebra of matrix valued fields  
on noncommutative space is thought to be the simple tensor product of 
constant matrix algebra and the Moyal-Weyl deformation.  
We propose scenarios in which the two become intertwined and 
inseparable.    Therefore the usual separation of 
ordinary or noncommutative space from the 
internal discrete space responsible for non-Abelian 
symmetry is really the exceptional case of 
an unified structure.  We call it 
{\em non-Abelian geometry}.  This general structure emerges when  
multiple D-branes are configured suitably in a flat but varying 
$B$ field background, 
or in the presence of non-Abelian gauge field background.  
It can also occur in connection with Taub-NUT geometry.  
We compute the deformed product of matrix valued functions using 
the lattice string quantum mechanical model developed earlier.  
The result is a new type of associative algebra defining 
non-Abelian geometry.
Possible supergravity dual is also discussed.

\Date{November 2000}
{\vfill\eject}
\ftno=0

\lref\CDS{A.~Connes, M.R.~Douglas and A.~Schwarz,
  {\it ``Noncommutative Geometry and Matrix Theory:
  Compactification on Tori,''} JHEP {\bf 9802} (1998) 003, hep-th/9711162.}

\lref\DH{M.~R.~Douglas and C.~Hull,
  {\it ``D-branes and the Noncommutative Torus,''}
  JHEP {\bf 9802} (1998) 008, hep-th/9711165.} 

\lref\SWNCG{N.~Seiberg and E.~Witten,
  {\it ``String Theory and Noncommutative Geometry,''} 
  JHEP {\bf 9909}, 032 (1999, hep-th/9908142.}

\lref\BerGan{A.~Bergman and O.~J.~Ganor,
  {\it ``Dipoles, Twists and Noncommutative Gauge Theory,''}
  hep-th/ 0008030.}

\lref\CDGR{S.~Chakravarty, K.~Dasgupta,
   O.~J.~Ganor and G.~Rajesh,
    {\it ``Pinned Branes and New Non-Lorentz Invariant Theories,''}
   Nucl. Phys. {\bf B587} (2000) 228, hep-th/ 0002175.}

\lref\HASI {A.~Hashimoto and N.~Itzhaki,
  {\it ``Non-commutative Yang-Mills and the AdS/CFT correspondence,''}
  Phys. Lett. {\bf B465} (1999) 142,
  hep-th/9907166.}

\lref\MALRUS{
  J.~M.~Maldacena and J.~G.~Russo,
  {\it ``Large N limit of non-commutative gauge theories,''}
  JHEP {\bf 9909}, 025 (1999), hep-th/9908134.}

\lref\ZYIN{Z.~Yin, 
   {\it ``A Note on Space Noncommutativity,''}
   Phys.\ Lett. {\bf B466} (1999) 234, hep-th/9908152.}

\lref\DGR {K.~Dasgupta, O.~J.~Ganor and G.~Rajesh,
  {\it ``Vector Deformations of $\cN = 4$ Yang-Mills Theory, Pinned Branes
 and Arched Strings,''} hep-th/0010072.}

\lref\DolNap {L.~Dolan and C.~Nappi, 
  {\it ``A Scaling Limit with Many Noncommutativity Parameters,''}
   hep-th/0009225.}

\lref\RT {R.~Tatar, {\it ``A Note on Noncommutative Field Theory and 
  Stability of Brane- Antibrane Systems,''}
 hep-th/0009213.}

\lref\WITSFT {E.~Witten, {\it ``Noncommutative Geometry and String Field
  Theory,''} Nucl. Phys. {\bf B268} (1986) 253.}

\lref\DGRunp {K.~Dasgupta, O.~J.~Ganor and G.~Rajesh, {\it unpublished}.}  

\lref\CHU{C.-S.~Chu and P.-M.~Ho, {\it ``Constrained Quantization of Open 
 String in Background $B$-Field and Noncommutative D-Brane,''}
Nucl. Phys. {\bf B568} (2000) 447, hep-th/9906192.}

\lref\KONTS {M.~Kontsevich, {\it ``Deformation Quantization of Poisson 
 Manifolds,''} q-alg/9709040.} 

\lref\CandF {A.~S. Cattaneo and  G.~Felder, {\it ``A Path Integral Approach 
 to Kontsevich Quantization Formula,''}
  Comm. Math. Phys. {\bf 212} (2000) 591, math.qa/9902090.} 

\lref\DGS{ K.~Dasgupta, G.~Rajesh and S.~Sethi, {\it ``M-Theory,
 Orientifolds and G-Flux,''} JHEP {\bf 9908} (1999) 023, hep-th/9908088.} 

\lref\BS{D.~Bigatti and L.~Susskind, 
{\it ``Magnetic fields, branes and noncommutative geometry,''}
Phys.\ Rev.\  {\bf D62}, 066004 (2000)
hep-th/9908056.}

\lref\KRSav{K.~Dasgupta, G.~Rajesh and S.~Sethi, {\it Work in Progress}.}

\lref\DouglasBE{
M.~R.~Douglas,
{\it ``D-branes on Calabi-Yau manifolds,''}
math.ag/0009209.}

\section {Introduction}

	This paper is devoted to the search and study of 
certain unusual and hitherto unknown 
facets of noncommutative space from string and 
field theories and quantum mechanics.  Introducing 
noncommutativity 
as a way of perturbing a known field theory has received 
much interest recently (see \refs{\CDS,\DH,\SWNCG} and the 
references therein),
 and hence we shall refrain from repeating 
the usual motivations and excuses for doing it.  Another  
reason lies in string theory itself.
The antisymmetric tensor field $B$ in the Neveu-Schwarz $-$ Neveu-Schwarz 
sector of string theory, while simpler than it cousins in the 
Ramond $-$ Ramond sector, is 
still shrouded in mystery and surprisingly resistant to 
an unified understanding.  One of its many features is its
relation to spatial noncommutativity.
Let us recall it briefly.

	Open strings interact by 
joining and splitting.  This lends naturally to the picture 
of a geometrical product of open string wave functionals 
that is clearly noncommutative.  One may formulate 
a field theory of open strings based on this noncommutative 
product the same way as conventional field theory is formulated 
on products of wave function fields\refs\WITSFT.  But the string 
wave functional 
is unwieldy and its product is enormously complex.  Noncommutativity 
certainly does not help.  To learn more we have to do with less.  
One way is to truncate the theory to a low energy effective theory 
of the small set of 
massless fields.  Another is to approximate the string by a minimal 
``lattice'' of two points.  This is especially well suited to mimicking   
the geometric product of open strings.  
It emerges from both approximations that, 
at least using some choice of variables, the natural 
product of wave function fields is the following 
noncommutative deformation of the usual one:
\eqn\StarProduct {
	(\Psi * \Phi) (x) = 
	 \exp\paren{\frac \imath 2 \ppa {{x'}^{\mu}} 
		\Omega^{\mu\nu} \ppa {{x''}^\nu} }
	\Psi(x') \Phi (x'') \Bigg{\vert}^{x' = x'' = x}.}
The parameter of noncommutativity $\Omega$ is expressed in 
terms of the spacetime metric\foot
	{Note that $G$ in this paper and in \refs\ZYIN is the
	same as the ``closed string metric'' $g$ in\refs\SWNCG, and 
	the noncommutativity parameter 
	$\Omega$ here is the same as $\Theta$ there.} 
$G$ and $B$ by 
\eqn\DefinitionOfOmega {
	\Omega = - (2 \pi \alpha')^2 G^{-1} B G^{-1}
	\paren {1 - (2 \pi \alpha')^2
	B G^{-1} B G^{-1}}^{-1}. }
It should be noted that noncommutativity is not a 
consequence of $B$ being nonvanishing or large.  It is 
intrinsic to the geometry of smooth string junctions 
that a canonical product exists for the functions 
on the space of open paths in the target space manifold 
with the appropriate boundary condition, and that product is noncommutative. 
The approximations mentioned above induces noncommutativity 
in the algebra of functions on the submanifold 
to which the end points 
are restricted, namely the D-brane.  The algebra actually becomes commutative 
in the limit of very large $B$!

	It is a glaring deficiency of the present understanding 
from string theory that one knew only how to deal with constant 
and flat $B$ field.  Introducing curvature for $B$ takes 
string theory away from the usual sigma model to a rather different 
realm, so understanding it fully seems to call for 
drastic conceptual advance.  On the other hand, 
varying but flat $B$ field 
should be accessible by the available technology 
but is hampered by technical difficulties.  For example, 
a formal construction of a noncommutative product using 
an arbitrary Poisson structure in place of the constant $\Omega$ has 
been given by Kontsevich\refs\KONTS.  
The construction made essential use of 
a degenerate limit of sigma model \refs\CandF.  But the result 
employs some very abstruse mathematics and its convergence properties 
essentially unknown.  
Behind the complication must lie some interesting and 
novel structures that needs to be deciphered.

	We propose, as a first step toward understanding such 
situation from string theory, 
probing it with multiple parallel D-branes configured in a way such 
that each D-brane only senses a constant but respectively 
different $B$ field.  On each D-brane the usual noncommutative 
algebra incorporates the effect of the locally constant $B$ field
without knowing that $B$ is actually varying.  The latter is 
revealed in the communication among different D-branes via  
the open strings that start and end on different D-branes.  
We study the wave functions associated with such ``cross'' strings 
and find that their product is deformed 
in a new and intriguing way that retains associativity.  
Along the same line of the reasoning as in \refs\SWNCG, 
one expects that it is in terms of this product that 
the effect of $B$ field is best 
described at the zero slope limit.
As D-branes are dynamical objects 
inclined to fluctuate, this picture 
is necessarily an idealization, describing the limit where the 
effect of such fluctuation is very small.  It would be worthwhile 
to study quantitatively the corrections due to such fluctuation.

	One can better appreciate the import of 
this new deformation of algebra by recalling another player.  
It is an essential feature of a
D-brane that it has a gauge symmetry and an associated gauge connection.  
Let us briefly recollect some of the well known facts relevant here.
In the simplest and most common circumstances, a single D-brane has 
an $U(1)$ symmetry, and a multiplet of $N$ D-branes on top of 
each other collectively have an $U(N)$ symmetry, with fields that 
are $N\times N$ matrices transforming in the adjoint representation 
of this $U(N)$.  
When the algebra of functions on the D-brane submanifold is deformed, 
so is the gauge symmetry.  For $U(1)$, the new Lie algebra is given 
by the commutator of the deformed product and is no longer trivial.  
For $U(N)$, the product of the matrix valued functions becomes 
\eqn\StarDeformedMatrixProduct {
	\left(M * N\right)^i_k (x) 
	= \sum_j \left( M^i_j * N^j_k \right) (x). }
And the deformed Lie algebra originates from the commutator of 
this new matrix algebra.  In this deformation 
the noncommutativity of spacetime 
and the non-Abelian property of multiple D-branes are 
simply and independently tensored together and do not affect each other, yet.

	It has long been known that the $U(1)$ (trace) part of 
the field strength $F$ always appear together   with $B$ 
as $(B-F_{U(1)})$.  A certain gauge symmetry actually connects 
the two.
Therefore $F$ also contributes to noncommutativity and appears 
in the 
expression for $\Omega$ by replacing $B$ with $(B-\tr F / N)$.  
What about the non-Abelian part  of $F$?  Let us consider 
a constant background for $F$, as varying $F$ is again too 
difficult.  For this constraint to make sense it has to be 
$U(N)$ covariant, i.e. it should be covariantly constant.  
Hence we also choose $F$ 
so that different spatial components of $F$ are 
in some Cartan subalgebra of $N\times  N$ matrices.  
By a choice of basis we can 
make them all diagonal.  Such background generically 
breaks $U(N)$ down to 
$U(1)^N$ and it is 
meaningful to talk about $N$ distinct 
branes, each with a constant field strength of its own unbroken $U(1)$.
This poses the same problem as the early configuration of 
multiple D-branes probing transversally varying $B$ field but 
with a different interpretation\foot {
	One difference is that in the configuration with varying 
	flat $B$ field, an open string stretching between two
	D-branes would have a mass offset proportional 
	to the separation between them.  It's possible to take a 
	special limit for the components of the close string metric along 
	the separation of the D-branes to make the offset vanish.  
	However, here we are only studying the kinematics 
	encoded in the algebra connecting all the the $(i,j)$ strings  
	and this offset is irrelevant.}. 
Now we consider a 
intrinsically non-Abelian deformation of the matrix product 
on the D-branes.  As it turns out, this new product is no 
longer the simple tensoring of the star product \eqr\StarProduct 
and the usual matrix algebra.  The noncommutative ``real'' 
space and the
non-Abelian internal space mix and become inseparable.  We call 
this {\it non-Abelian geometry}, and give a general formulation and 
the underlying philosophy at the end of section 3.

In this work, we have found a large class of  examples of this new geometry 
by considering non-trivial D-branes configurations with non-Abelian field 
content and/or under the influence of non-flat $B$ field.  The concrete
form of the product are derived from a lattice approximation of string theory 
in section 2 and 3, and presented here.  There are different ways to express 
the product, corresponding to different choices of operator ordering.  
With the ``symmetric'' ordering defined and used throughout
section 2, the geometry is defined by an algebra with the following
product.
\eqn\GeneralSymmetric{ 
 (\Psi * \Phi)^i_j(x) \equiv \sum_l
\exp\paren{ \frac \imath 2\ppa{{x'}^{\mu}} 
	\Omega^{\mu\nu}_{i l; l k}\ppa{{x''}^{\nu}}}
          \Psi^i_l(x')\Phi^l_k(x'')\Bigg{\vert}^{x'=S^{ik}_{ij}x}
_{x''=S^{ik}_{jk}
          x} . }
with $S$ and $\Omega$ satisfying (section 2.2)
\eqn\GeneralCocycleCondition{\eqalign {
	& ~~S_{j_1 j_2}^{j_3 j_4} 
	= S_{j_1 j_2}^{k_1 k_2} S_{k_1 k_2}^{j_3 j_4};\cr
	& ~~S_{j_1 j_2}^{j_3 j_4} \Omega_{j_3 j_4; j_5 j_6} =
	\Omega_{j_1 j_2; j_5 j_6} =
	\Omega_{j_1 j_2; j_3 j_4} {S_{j_5 j_6}^{j_3 j_4}}^{\top}.}}
Here no summation over Latin (Yang-Mills) indices takes place, but summation 
over repeated suppressed Greek (space-time) indices does take place.
With the split ordering introduced in section 3, the product 
takes the form 
\eqn\GeneralSplit{
	\paren {\Psi \times \Phi}^i_j (e^a, e^A)
	\equiv  \sum_{k,l} \Psi^i_k(e^a, M^A) 
	\exp\paren{\imath \ppa {M^A} {\hat{\bf{\Omega}}}^{Aa} \ppa {M^a}}_{kl}
		\Phi^l_j(M^a, e^A) \Bigg{\vert}^{M^A = 0 = M^a},	}
Here ${\hat{\bf{\Omega}}}^{Aa}$ is a matrix with Latin indices $k$ and $l$ 
the exponential is the usual exponential of a matrix.  In this 
paper we shall derive \eqr\GeneralSymmetric and \eqr\GeneralSplit 
from certain physical backgrounds in string theory, so they appear 
in concrete and specific context with  
motivation from string theory.  However, we should emphasize here that 
with the forms of the products now known, we can and should 
consider them independently 
and abstractly as examples of non-Abelian geometry, and look for 
their appearance in other contexts as well.  
In deriving \eqr\GeneralSymmetric the parameters $S$ and $\Omega$ takes on 
values determined by the physical background in our setup.  
However, \eqr\GeneralSymmetric stands as a valid 
definition of an associative product as long as
\eqr\GeneralCocycleCondition holds.  Similarly, we have derived
\eqr\GeneralSplit in section 3 for the case of $SU(2)$, but these forms of
product apply more generally for arbitrary ${\hat{\bf{\Omega}}}_{kl}^{Aa}$.  
Unlike the symmetric ordering, here $\hat {\bf \Omega}$ 
is already ``gauge fixed'' 
and associativity imposes no constraint on them.
The two form should be related to
each other by a change of ordering.  This is clear in the examples
discussed in this paper, though we have not worked out the general and
explicit transformation relating the two in this paper.  Finally we note that even in
such general forms, they only represent a certain class of non-Abelian
geometry.  The explicit forms of non-Abelian geometry in its full
generality is a very interesting problem still under investigation.

At this point one may well consider other approaches to generalizing D-brane 
geometry.  One very interesting approach in recent time has been the 
efforts to study the geometry of D-branes with vector bundles in Calabi-Yau
manifolds (\refs\DouglasBE and the references therein).  There one 
takes the D-brane wrapping supersymmetric cycles in Calabi-Yau manifolds 
and the vector bundle on the D-branes as a whole and study their properties 
in relation to target space supersymmetry and mirror symmetry.  It would 
be very relevant to fully reconcile these two facets of D-branes:
the vector bundle aspects of D-branes steeped in conventional commutative 
geometry as one, and the non-commutative geometry 
that the open string fields see as the other.   However, non-commutativity 
introduces such tremendous technical challenges in curve space that 
novel and powerful methods and concepts seems necessary to tackle it.
This paper provides one possibility in dealing with curvature in the 
antisymmetric tensor $B$ field (section 2.2).  It might prove 
helpful in dealing 
with curvature in the metric through various correspondences such as 
T-duality, which relates the metric and the $B$-field.

Here is an outline of the paper.  
The non-Abelian noncommutative product is explicitly constructed 
in section two.  It turns out a two point lattice approximation 
to quantum mechanics is perfectly suited for this purpose.
A systematic methodology of computing 
product was developed in \refs\ZYIN.  
We review and elaborate it in section 2.1.  In section 2.2 we apply it 
to the most general case of non-Abelian noncommutativity and obtain the main
result of the paper (eq. 2.55).  
In section 3 we then turn to the specific case of 
the deformation parameter being in the adjoint of $SU(2)$ and use 
a variation of the method presented in section two.
The result is 
a surprisingly compact and highly suggestive form of the new 
product (eq. 3.16).
The situation of multiple noncommutativity 
parameters also makes an appearance 
in connection with Taub-NUT geometry\foot{For branes near a conifold,
non-Abelian
noncommutativity makes an appearance in the fractional brane setup
\refs\RT.}.  
We shall discuss in section four how 
a whole class of 
Lorentz non-invariant
theories governed by the $B$ field dynamics 
can be studied in an unified way. 
We conclude with a discussion on
the possible 
gravity dual for the system we study as well 
as some other related issues.

\section {Construction of the non-Abelian noncommutative product}

\subsection {Review and elaboration}

\subsubsection {The origin of noncommutativity}

	A classic and salient feature of string theory is 
its geometric appeal.  For example, strings interact
by smoothly joining and splitting.  In conventional field 
theories, one can visualize 
an interaction of particles  
as a vertex of intersection by 
propagators in a Feynman diagram.  The well known rule 
from perturbation theory states that each term in the 
interaction Lagrangian gives rise to a distinct kind of such 
vertex.  Interaction at a point corresponds to product 
of fields at the same point.  The rule of string theory 
perturbation is entirely analogous.  However, the algebra 
of the product, besides being obviously much more complicated, 
has a new twist.  Consider the joining of two or more 
open strings into 
one.  It should be apparent that this process is {\em not} 
commutative though still clearly associative.  
The multiplication between the wave functionals of the open string, 
also known as the open string fields, share the same 
property\foot 
	{This observation was made clear in \refs\WITSFT, 
	where one can also find relevant graphic illustrations.}.

	Intuitively, the product seems easy to define.
Let $\Psi[\gamma]$ and $\Phi[\gamma]$ 
be two string wave functionals, 
where the argument $\gamma$ is an open path in the target space 
with the proper boundary conditions.  
The geometric product defined above can be written as 
\eqn\OpenStringJoiningProduct {
	\left( \Psi \cup \Phi \right) [\gamma] 
	= \myI {D[\gamma_1] D[\gamma_2]}  \,\,
	\delta [\gamma=\gamma_1\uplus\gamma_2]
	\,\, \Psi [\gamma_1] \, \Phi [\gamma_2] ,   }
The operation $\uplus$ is just the 
geometric process of ``joining'' defined above, with a refining 
sensitivity to sign and 
orientation so that a segment that backtracks itself also erases 
itself.  This ``definition'' 
manifests noncommutativity\foot 
	{Even though we have made no mention of $B$!}
and associativity, but it is also 
horribly divergent and ill defined.  One can remedy this with an elaborate 
procedure \refs\WITSFT but there is an alternative way 
to make sense of this product, if one is willing to forgo 
the bulk of the data encoded in the string field in exchange 
for a better understanding of what remains.

	Before we do this first recall that 
the standard string 
action is 
\eqn\OpenStringAction { \eqalign {
	S &= \frac 1 {4 \pi \alpha'}
	\int_{\Sigma} \! d^2 \sigma \,
	   \brak { G_{\mu\nu} \paren{{\dot X}^\mu {\dot X}^\nu
			- { X'}^\mu {X'}^\nu }}	\cr
	& +  \int_{\pa_2 \Sigma} \! d\tau \,
		\cA_\mu(X) \dot X^\mu
	-  \int_{\pa_1 \Sigma} \! d\tau \,
		\cA_\mu(X) \dot X^\mu. } }
Here the subscripts ``$2$'' and ``$1$'' on $\pa \Sigma$ 
label the ``left'' and ``right'' boundaries of the open string worldsheet.
$G$ is the background closed string metric and assumed to be 
constant.  
Usually there would also be a term 
$\frac 1 {4 \pi \alpha'} B_{\mu\nu} \paren{{\dot X}^\mu {X'}^\nu
			- { X'}^\mu   {\dot X}^\nu } $
in the action.  However we would only be dealing with flat $B$ field, 
and in $R^D$ flat 
$B$ is exact and equal to $dA'$ for some $A'$.  We henceforth 
include $-A'$ implicitly in $\cA$ so that $d\cA = F - B \equiv \cF$.  

Let us be careful with boundary conditions from now on.  
To solve the equation of motion we need to impose 
one for each boundary component.  We want the two ends of the strings to 
move only within two possibly distinct but parallel D-branes of 
the same dimensions.   
For 
the purpose at hand, we will only be concerned with coordinates 
that parameterize the D-branes' worldvolume under the influence of 
a nondegenerate $\cF$ and ignore from now on 
all the other coordinates, including those along which the 
D-branes separate.  We shall only consider situations for which 
this space is $R^D$.  The boundary condition for the 
relevant coordinate fields is 
\eqn\OpenStringBoundaryCondition	{
	\frac 1 {2 \pi \alpha'} G_{\mu\nu} {X'}^\nu
		= \cF_{\mu \nu} {\dot X}^\nu.	}
For the problem to be tractable using the method in this paper, 
we also require $\cF$ to be constant.  Note that since $\cF$ is 
evaluated only at the ends of the strings, confined to the D-branes, 
this requirement only enforces the constancy of 
the pull-back of $B$ to the D-branes.  $B$ may vary in directions 
transverse to the D-branes, or have components not entirely 
parallel to the D-branes that vary.  Indeed the flatness of $B$ 
correlate the last two kinds of variations.  In this subsection 
let us consider the case of a single D-brane, so there is 
only one constant $\cF$.  We will return to the general case in 
the next subsection.

	Now we approximate the spatial extent of the
open string by the coarsest ``lattice'' of two points, namely 
the two ends, labeled by $1$ and $2$.  Let the width of
the string be $2 / \omega$.  The
action \eqr {\OpenStringAction} is approximated by\foot 
	{A similar but different model appeared in \refs\BS.}
\eqn\DipoleAction	{ \eqalign {
	S &=
	\myI {d\tau}
	   \brak { \frac 1 {4 \pi \omega \alpha'}
	\paren{ {\dot X}_1^2 + {\dot X_2}^2
			- \frac {\omega^2} 2 ( X_2 - X_1 )^2 }}	\cr
	& +  \myI {d\tau}
		\paren {A_\mu(X_2) \dot X_2^\mu
			-  A_\mu(X_1) \dot X_1^\mu}.	}}
We shall call this system {\em lattice string quantum mechanics} (LSQM).
The boundary conditions now become\refs\ZYIN 
\eqn\DipoleConstraint	{	\eqalign {
	D^1_\mu &\equiv [G (X_1 - X_2)]_\mu
	   +  \frac {4\pi\alpha'}  \omega
	  [\cF \dot X_1]_\mu \sim 0; \cr
	D^2_\mu &\equiv [G (X_2 - X_1)]_\mu
	   -  \frac {4\pi\alpha'}  \omega
	   [\cF \dot X_2]_\mu \sim 0. } }
The result of canonical quantization with constraints is \foot 
	{In comparing the results summarized here with \refs\ZYIN 
	~one should note that $D^i$ defined here is equal to 
	$G_{\mu\nu}C_i^\nu$ in \refs\ZYIN, and that there is 
	a typographical error of a missing $(-1)$ in front the 
	expression on the third line in (eq.~2.12) of \refs\ZYIN, and 
	another $(-1)$ on the exponent of the 
	second 	parenthesis in the same expression.}
\eqn\CommutatorOnBoundary	{ 	\eqalign {
	\comm {X_2^\mu} {X_2^\nu} =&   
		-\imath (2 \pi \alpha')^2 \brak {G^{-1}\cF G^{-1}
	\paren {1 - (2 \pi \alpha')^2
		\cF G^{-1} \cF G^{-1}}^{-1} }^{\mu\nu} \cr
		\equiv& ~~\imath \Omega^{\mu\nu} 
		= -\comm {X_1^\mu} {X_1^\nu}; \cr
	\comm {X_1^\mu} {X_2^\nu} =&~ 0.	}	}
These are precisely the commutation relations for the ends 
of the string found in \refs\CHU.

\subsubsection {Matrix, Chan-Paton Factor, and Noncommutative Product}

	We are now only one step from 
the deformed product \StarProduct.  It is here that 
the LSQM approach distinguishes itself for its 
conceptual and technical advantage.  Now that the entire 
continuum of the open string is distilled down to two points, 
the above mentioned $\cup$ joining of two oriented paths 
into one  
reduces to the 
merging of two ordered pairs of points, with the second (end) 
of the first pair coinciding with and ``cancelling'' 
the first (start) of the second 
pair: $(x_1, m) \uplus (m, x_2) = (x_1, x_2)$.  This induces 
a product $*$ of two wave functions of the lattice string, 
entirely analogous to \OpenStringJoiningProduct:
\eqn\LatticeStringJoingingProduct {
	\left( \Psi * \Phi \right) (x_1, x_2) 
	= \myI {dm_1 dm_2}~ \delta (m_1 - m_2) 
	\Psi (x_1, m_2) \Phi (m_1, x_2).	}

	If this seems reminiscent of ordinary matrix product, 
it is no illusion.  One can think of an index  
on a matrix as a coordinate parameterizing some discrete space\foot{The 
precise connection is the discrete ``fuzzy'' torus explained later in
footnote (13) in section 3.}.   
Since a matrix carries two indices 
it is the wave function of a lattice string moving 
on this discrete space and its $*$ product would simply be 
the standard matrix multiplication.  
Distinguishing between the contravariant and covariant indices 
corresponding to distinguishing the two ends 
of the (lattice) string by a choice of orientation.  
On the other hand, 
attaching discrete indices to string ends is none other than 
introducing Chan-Paton factors.  In this light, the noncommutativity 
of open string field and of non-Abelian gauge symmetry are not just 
similar in their failure to commute but have a 
shared geometric origin and interpretation!

	Now we return to {\em LSQM} 
(lattice string quantum mechanics).  Its 
salient feature, reviewed shortly, is the 
truncation of the noncommutative string field algebra down to 
a closed noncommutative algebra of (wave) functions on the target space. 
The latter is something much simpler and easier to study 
than the full open string algebra and still carries 
nontrivial information, especially the effects of the $B$ field.
The known noncommutative algebra found this way is a deformation 
of the ``classical'' commutative algebra of functions.  
It modifies the $U(1)$ gauge symmetry of a single D-brane experiencing 
this $B$ field into a deformed one corresponding to the group 
of unitary transformations in a certain Hilbert space.  
When multiple D-branes are present so that $U(1)$ is replaced by 
the non-Abelian $U(N)$, the $U(N)$ group as well as the $N\times N$ 
matrix algebra is also modified.
The new algebra is just the tensor product of the matrix algebra 
and the deformed noncommutative algebra of the scalar functions. No 
essential difference in the noncommutativity of the space is 
introduced by having non-Abelian gauge symmetry.  
This then begs the question:  is there some other deformation in which the 
discrete internal space can  
become fully entangled with the (noncommutative) ``real'' 
space so that it is impossible to separate them.  
The answer, we shall propose, is {\it yes}.  The condition, we shall show,
is that the background parameter for noncommutativity is in
an appropriate sense 
{\it non-Abelian}.  This can be due to a non-$U(1)$ background 
for the gauge field or a varying flat $B$ configured in the manner 
prescribed  
above. 

\subsubsection {Defining the product}

	The \eqr\LatticeStringJoingingProduct  almost entirely
defines the rule for making product.  We still have the trivial 
freedom of changing the overall normalization by a constant 
factor, which we will fix later.  Yet that equation would seem
to be applicable to functions on the square of $R^D$ 
rather than $R^D$ itself.  Fortunately, 
\eqr\CommutatorOnBoundary says that 
although we start from $2D$ canonical coordinates in the LSQM, constraint 
\eqr\DipoleConstraint reduces the size of a complete set of commuting 
observables to only $D$, the right number 
for a wave function to be defined on $R^D$ itself.  
At each 
of the two ends, there are only $D/2$ commuting observables.  
Let us make some choice and call them $E^a_1$ and $E^a_2$, 
$a = 1\ldots D/2$, where the subscript labels  
boundary components.  Together they form a \CSCO. 
We shall call it the ``$aa$'' representation which diagonalizes 
simultaneously $E^a_1$ and $E^a_2$  with 
eigenvalues $e^a_1$ and $e^a_2$ respectively.
For wave functions $\Psi_{aa}(e_1^a, e_2^a)$ 
and $\Phi_{aa}(e_1^a, e_2^a)$
in this representation, the 
adaptation of \eqr\LatticeStringJoingingProduct is immediate 
and obvious: 
\eqn\aaProduct {
	\paren {\Psi * \Phi}_{aa}(e_1^a, e_2^a)
	\propto \myI {dm_1^a dm_2^a} ~ \delta (m_1 - m_2)
	\Psi_{aa}(e_1^a, m_2^a)  \Phi_{aa}(m_1^a, e_2^a). }
The proportionality sign here signifies that we have yet to 
specify the overall normalization, which scales the right 
hand side of \eqr\aaProduct by a constant factor.  
We will fix it later by relating to the usual commutative 
product.

	This product is natural  also 
from the point of view of the LSQM.  $X_1$ and $X_2$ commute 
with each other.  Therefore  
the left and right ends decouple and the Hilbert 
space for the LSQM is the tensor product of the Hilbert spaces 
$\cH_1$ and $\cH_2$ of two ends 
respectively\foot 
	{Subtlety might arise for other topology 
	but at least for $R^D$ this factorization holds.}.
Furthermore, the operator algebra in $\cH_1$ and $\cH_2$ are 
generated by the same set of observables $X^\mu$, but from 
\eqr\CommutatorOnBoundary their commutator is exactly 
opposite in sign.  
This canonically correlates 
 them as complex conjugate pair of 
representations of the same operator algebra.  To see this, 
choose a basis of $R^D$ so that $\Omega$ is brought to 
the canonical form:
\eqn\DefinitionOfJ {
	J = \left( \matrix {
	0 &	\id \cr
	-\id & 	0  \cr} \right). }
Let  $a, b, \ldots$ enumerate  the first $D/2$ coordinates 
and $A, B, \ldots$ the rest. Then \eqr\DefinitionOfJ can be written more
compactly as:
\eqn\DefinitionOfe { \eqalign {
	J^{aA} &= \delta^{a+{D\over 2},A} = - J^{Aa}, \cr
	J^{ab} &= 0 = J^{AB} .\cr  }}
In the $aa$ representation the $E^a$'s are simultaneously diagonalized 
while the $E^A$'s are implemented as differentiations:
\eqn\DefinitionOfeA	{	\eqalign	{
	E_1^A &= - \imath J^{aA} \left(\ppa {e_1^a} 
		+ \imath \ppa {a} \alpha_1(e_1^a)\right), \cr
	E_2^A &= \imath J^{aA} \left(\ppa {e_2^a} 
		+ \imath \ppa {a} \alpha_2 (e_2^a)\right). }}
Here $\alpha_1$ and $\alpha_2$ are just the usual phase ambiguity 
in canonical quantization.  We can naturally identify $e_1$ and $e_2$ 
by identifying the wave functions in $\cH_2$ to $\cH_1$ after 
complex conjugation, 
and requiring $\alpha_1 = - \alpha_2$.  
Thus a ket in $\cH_2$ is a bra in $\cH_1$ and vice versa.  
The product \eqr\aaProduct can be rewritten as 
\eqn\BraKetProduct	{
	(\ket {\alpha} \otimes \bra {\beta}) 
	* (\ket {\theta} \otimes \bra {\rho})
	\propto
        (\bra \beta \ket \theta) (\ket \alpha \otimes \bra \rho).	}

	Although this product 
has manifest noncommutativity and associativity, the 
wave functions are not functions on the target space and there is 
no parameter visible that controls the noncommutativity.
This is a fitting time to remember that an associative 
algebra has both additive and multiplicative structures but 
\eqr\aaProduct defines only the latter.  We want our algebra to be 
a deformation, in its multiplicative structure, of the algebra 
of functions on $R^D$, so it should be identified with 
the set of functions on $R^D$ as a vector space.  Wave functions 
in the $aa$ representation clearly does not suit this purpose.  
We need to find an ``$aA$'' representation in which a set of $D$ 
observables that can pass as coordinates on $R^D$ are simultaneously 
diagonalized.  That is tantamount to requiring 
the action of translation generator $P$ on them should 
be what is expected of $R^D$ coordinates.  We call them 
{\em geometric observables}.

	For example, in the LSQM above $P = \Omega^{-1}(X_1 - X_2)$.  
A particularly symmetric choice for the geometric observables is 
simply the center of mass coordinates $X_c^\mu$ of the lattice string system: 
\eqn\SymmetricXc	{
	X_c^\mu = {1\over 2}(X_1^\mu + X_2^\mu)  }
such that 
\eqn\ActionOfTranslation	{
	\comm {X^\mu_c} {P_\nu} = \imath \delta^\mu_\nu.}
We can rewrite the product \eqr\aaProduct in terms of 
functions of the eigenvalue $x$ of $X_c$ by using the change of basis 
function\foot 
	{ $\ket {x}$ is an eigenstate of $X_c$ and $\ket {e_1^a,e_2^a}$
(shorthand for $\ket {e_1^a} \otimes \bra {e_2^a}$) that for $E_1^a$ and 
$E_2^a$. 
    We have also made a convenient choice for the phase 
	for the basis wave function of these representations.}
\eqn\SimplestCaseOfChangeOfBasis	{
	\bra {x} \ket {e_1^a, e_2^a}
	= \delta \left(x^a - {1\over 2} (e_1 + e_2)^a  \right)
	\exp {\left( - \frac \imath 4 x^A J_{Aa}
		(e_2 - e_1)^a \right)}	}
and find that, in terms of $\Omega$, 
\eqr {\aaProduct} is explicitly given by
\eqn\StarProduct	{
	(\Psi * \Phi) (x) =  \exp {\paren {\frac \imath 2
		\Omega^{\mu\nu} \frac \pa {\pa{x'^{\mu}}}
		 \frac \pa {\pa{x''^{\nu}} }}}
	\Psi(x') \Phi(x'') \Bigg{\vert}^{x'=x''=x} .	}
Here we have fixed the overall normalization mentioned before by 
requiring it to reproduce the usual commutative product when 
$\Omega = 0$.

\subsection{ Non-Abelian Deformation}

	Now we come to the main task of this paper and 
consider the 
possibility of more than one noncommutativity parameter.  
For each of such parameter we can define the $\ast$ product 
above and have a distinct algebra.  
Let us assign labels ranging from $1$ to $N$ to this 
group of parameters $\Omega_i$.  We denote elements of 
the $i$-th algebra by functions labeled such as $\Psi^i_i$, 
satisfying 
\eqn\noname	{
	(\Psi^i_i * \Phi^i_i) (x) = 
	 \exp {\paren {\frac \imath 2
		\Omega^{\mu\nu} \frac \pa {\pa{x'^{\mu}}}
		 \frac \pa {\pa{x''^{\nu}} }}}
	\Psi^i_i(x') \Phi^i_i(x'') \Bigg{\vert}^{x'' = x' = x}.	}
The reductionist view 
of what we want to do is to find a way to glue these algebras 
together cogently into one unified algebra.  For that we now 
return to string theory for intuition.

	In string theory the above situation can arise 
in a configuration of $N$ D-branes with 
different but constant $\cF$ on each of them.  
{}From the discussion of the last subsection, this can happen 
in an arbitrary combination of two scenarios.
The first, already explained before, is 
a background of flat but varying $B$ field configured in such a way 
 that (only) the pull-back of $B$ to each D-brane is constant 
through it.  The second scenario is a background gauge field that is 
constant but breaks the $U(N)$ gauge symmetry.   It is not in general 
meaningful to talk about constant non-Abelian 
curvature because it would normally not satisfy the equation of 
motion or the Bianchi identity, but if all the spatial components of 
the curvatures are in some Cartan subalgebra than everything is fine.  
For $U(N)$ this amounts to being able to diagonalize all spatial 
components of the curvature as $N\times N$ matrices.  This breaks 
$U(N)$ down to $U(1)^N$ and gives the interpretation of $N$ 
D-branes each with a distinct and constant $U(1)$ background. The $(i,i)$  
strings on each of the D-branes are now complemented by $(i,j)$ strings,
which start on the $i$-th brane 
and end on the $j$-th D-brane\foot
	{In \refs\SWNCG  a system
	of D0-D4 was studied with the $0-4$ strings having mixed boundary
	conditions. These strings complement the $0-0$ strings and the $4-4$
	strings to produce a bigger algebra.}.

	Consider wave functions $\Psi^i_j$ 
in the lattice string quantum mechanics approximating to the $(i,j)$ 
string.  The Hilbert space is a tensor product 
of $\cH_i \otimes \cH_j^*$ and the product rule of the whole 
algebra is generated by 
\eqn\NonAbelianBraKetProduct	{
	(\ket {\alpha}_i \otimes \bra {\beta}_j) 
	* (\ket {\theta}_j \otimes \bra {\rho}_k)
	= (\bra {\beta}_j \ket {\theta}_j)
	  (\ket {\alpha}_i \otimes \bra {\rho}_k).  }
Written in terms of matrix valued functions $\Psi$ and $\Phi$ 
on $R^D$, this is 
\eqn\CrossProduct {
	\paren {\Psi \times \Phi}^i_k (x) \equiv 
	\sum_j \paren {\Psi^i_j *_{ijk} \Phi^j_k} (x).	}
Note that in \eqr\NonAbelianBraKetProduct the product seems to 
depend only on $j$, but one has to write the final form 
\CrossProduct in the $aA$ representation.  In general that 
would mean $*_{ijk}$ depends on all three indices.  
Our goal is to calculate $*_{ijk}$.

\subsubsection {Preparation and Notations}

	Again each brane is labeled by index $i, j, \ldots$.
let us denote the $\cF$ on the $i$-th D-brane 
by $\cF^i$.  One repeats the same procedure of constrained quantization.  
This time one finds the Poisson brackets of 
the constraint $D$'s are 
\eqn\PoissonBracketOfConstraint	{	\eqalign {
	\PB {D^1_\mu} {D^1_\nu} &= - {4 (2 \pi \alpha')^2} 
			\brak {\cF_i \paren{1
		- (2 \pi \alpha')^2 \cF_i G^{-1} \cF_i G^{-1}}}_{\mu\nu}, \cr
	\PB {D^2_\mu} {D^2_\nu} &=
		~~ {4 (2 \pi \alpha')^2} 
			\brak {\cF_j \paren{1
		- (2 \pi \alpha')^2 \cF_j G^{-1} \cF_jG^{-1}}}_{\mu\nu}, \cr
	\PB {D^1_\mu} {D^2_\nu} &= ~~ 
	{2 (2\pi\alpha')^2} \brak{\cF_i-\cF_j}_{\mu\nu}. } }
	
	For $i\neq j$, $D^1$ and $D^2$ no longer commute.  
This would translate 
to $X_1$ and $X_2$ not commuting with each other and would 
impede the program we have developed for constructing the 
product.  However, we can take the zero slope limit employed 
in \refs\SWNCG, in which $\alpha' \to 0$ while 
$\cF$ and $(2\pi \alpha')^2 G^{-1}$ remain finite.

	After taking the limit, one finds that 
\eqn\BoundaryCommutationRelation {\eqalign {
	\comm {X_1^\mu} {X_1^\nu} &= -\imath \Omega_i^{\mu\nu}, \cr
	\comm {X_2^\mu} {X_2^\nu} &=~~ \imath \Omega_j^{\mu\nu}, \cr
	\comm {X_1^\mu} {X_2^\nu} &=~ 0,}}
where 
\eqn\DefinitionOfOmega	{
	\Omega_i = (\cF^i)^{-1}. }

We can always, through a congruence transformation, turn an 
$\cF$ into the following canonical form:
\eqn\CanonicalFormForF {
	\Omega_i = T_i J T_i^\top  ,}
where
\eqn\DefinitionOfJ {
	J = \left( \matrix {
	0 &	\id \cr
	-\id & 	0  \cr} \right). }

It shall become convenient to use the following symbols:
\eqn\TwoJMSymbol{\eqalign{
(1) & ~~	U_{ij} \equiv T_i T_j^{-1}  ;\cr
(2) & ~~	M^{j_1 j_2} \equiv T_{j_1}^{-1} + T_{j_2}^{-1};\cr
(3) & ~~	\cF^{j_1 j_2; j_3 j_4} \equiv
	- \frac {(M^{j_1 j_2})^\top J {M^{j_3 j_4}}} 4; \cr
(4) & ~~	\Omega_{j_1 j_2; j_3 j_4} \equiv 
	\left( \cF^{j_3 j_4; j_1 j_2} \right)^{-1} ;\cr
(5) & ~~	S_{j_1 j_2}^{j_3 j_4} \equiv 
	(M^{j_1 j_2})^{-1} M^{j_3 j_4} }}
They are related to each other and $\Omega_j$ by 
\eqn\RelationOfSOmegaSymbol {\eqalign {
	(1) & ~~{\cF^{j_1 j_2; j_3 j_4}}^\top = - \cF^{j_3 j_4; j_1 j_2}; \cr
	(2) & ~~\cF^j = \cF^{j j; j j}; \cr
	(3) & ~~\Omega_j = \Omega_{j j; j j}; \cr
	(4) & ~~S_{j_1 j_2}^{j_3 j_4} = \Omega_{j_1 j_2; k_1 k_2} 
		\cF^{k_1 k_2; j_3 j_4} 
	= S_{j_1 j_2}^{k_1 k_2} S_{k_1 k_2}^{j_3 j_4};\cr
	(5) & ~~S_{j_1 j_2}^{j_3 j_4} \Omega_{j_3 j_4; j_5 j_6} =
	\Omega_{j_1 j_2; j_5 j_6} =
	\Omega_{j_1 j_2; j_3 j_4} {S_{j_5 j_6}^{j_3 j_4}}^{\top}.}}
Note that because we will deal with a plethora of 
indices we shall suppress  spatial indices unless 
doing so will cause confusion. Repeated gauge indices 
$i,j$ are not summed over
unless stated otherwise explicitly, but repeated spatial indices 
are always summed over implicitly.
The situation should be obvious from the context.  Coordinates are arranged 
into column vector, or row vectors after transposition.

\subsubsection {In search of a center}

	In this subsection we figure out the geometric observables 
for the $(i,j)$ dipole.  That is, we want operators $X_c^\mu$ such that 
\eqn\DefinitionOfCenterCoordinates {
	\comm {X_c^\mu} {P_\nu} = \imath \delta^\mu_\nu,}
where the translation generator $P_\mu$ for the dipole system described by 
\eqr\DipoleAction is 
\eqn\TranslationGenerator {
	P = \cF^i X_1 - \cF^j X_2 }
in the zero slope limit taken earlier.
$P$ satisfies the property
\eqn\TranslationDoesNotCommute {
	\comm {P_\mu} {P_\nu} = -\imath (\cF^i_{\mu\nu} - \cF^j_{\mu\nu}) 
	\equiv -\imath \Delta_{\mu\nu}.}
Therefore they are like covariant 
derivatives and we require them to be as such:
\eqn\PIsCovariantDerivative {
	P \equiv \Pi + \tilde A,}
where
\eqn\DefinitionOfPi {
	\Pi_\mu = - \imath \frac \pa {\pa X_c^\mu} ,}
and 
\eqn\DefinitionOfTildeA {
	\pa_\mu {\tilde A}_\nu - \pa_\nu {\tilde A}_\mu
	= \Delta_{\mu\nu}.}
The definition of ${\tilde A}$ suffers the usual phase ambiguity 
and we choose a {\em linear} gauge
\eqn\GaugeChoiceForP {
	\tilde A = - {1\over 2}(\Delta + \cTheta )X_c ,}
where $\cTheta$ is a symmetric matrix and pure gauge.  
It will be fixed later for convenience.

	There are an infinite number of choices for operators 
satisfying \eqr\DefinitionOfCenterCoordinates.  
Let us for now look for one as close to the center of mass 
\eqn\HalfAndHalf {
	X_{1/2} = {1\over 2}(X_1 + X_2) }
as possible.  Alas
\eqn\BadNewsForXhalf {
	\comm {X^\mu_{1/2}} {X^\nu_{1/2}} 
	= - \frac \imath 4
		(\Omega^{\mu\nu}_i - \Omega^{\mu\nu}_j) 
	\equiv \imath \nabla^{\mu\nu},}
so $X_{1/2}$ itself does not suffice.  
Therefore we define $X_c$ indirectly 
so that  
\eqn\DefinitionForXhalf {
	X_{1/2} = \Lambda X_{c} + \Gamma \Pi  .}
Then 
$\Lambda$ and $\Gamma$ can be found by substituting 
\eqr\DefinitionOfPi and \eqr\DefinitionForXhalf into 
the commutation relation known for $X_{1/2}$ and $P$:
\eqn\PiIsConjugateToX {
	\imath \delta^\mu_\nu = \comm {X_{1/2}^\mu} {P_\nu} 
	= \imath \left(\Lambda - 
		{1\over 2}\Gamma \left(\Delta - \cTheta \right) \right) } 
and
\eqn\XandXDoesNotCommute {
	\imath \nabla^{\mu\nu} = \comm {X_{1/2}^\mu} {X_{1/2}^\nu} = 
	\imath (\Lambda \Gamma^\top - \Gamma \Lambda^\top) 
	 .}
It thus follows that $\Lambda$ is related to $\Gamma$ by 
\eqn\LambdaIsRelatedToGamma {
	\Lambda = \id + {1\over 2}\Gamma \left(\Delta - \cTheta \right)} 
So
\eqn\ConditionOnLambda {
	\Gamma^\top - \Gamma + \Gamma \Delta \Gamma^\top = \nabla. }
$\Gamma$ is thus related to a matrix $\gamma$ 
satisfying
\eqn\DefinitionOfgamma {
	\gamma \Delta \gamma^\top = \Delta - \Delta \nabla \Delta} 
through 
\eqn\RelationOfGammaWithgamma {
	\gamma = 1 + \Delta \Gamma.}

One can then solve for $X_c$ and find that
\eqn\SolutionForX {
	X_c = \left(1 + \Gamma \Delta \right)^{-1} 
	\left( X_{1/2} - \Gamma P \right)
	= \left(1 + \Gamma \Delta \right)^{-1} 
	\left( \left(\half + \Gamma \cF^i \right) X_1 
	+ \left( \half - \Gamma \cF^j \right) X_2 \right)  .}

\subsubsection {Solving for $\Lambda_{ij}$}

The matrix $U_{ij}$ defined in \eqr\TwoJMSymbol 
satisfy 
\eqn\TransformationBetweenOmega {
	\Omega_i = U_{ij} \Omega_j U^T_{ij}. }
as well as the cocycle condition
\eqn\CocycleCondition {
	U_{ij} U_{jk} = U_{ik}  .}
{}From the requirement that the 
$X_c$'s commute among themselves it follows that 
$$(1+2\Gamma \cF^i)^{-1} (1-2 \Gamma \cF^j)$$ 
also satisfy the condition 
\eqr\TransformationBetweenOmega.  We use 
this to find a solution for $\Gamma_{ij}$ in terms of 
$U_{ij}$:
\eqn\GammaInTermsOfU {
	\Gamma = \half \left( 1 - U_{ij} \right)
	\left( \cF^i U_{ij} + \cF^j \right)^{-1}} 
so that 
\eqn\UIntermsofLambda {
	U_{ij} = (1+2\Gamma \cF^i)^{-1} 
	(1-2 \Gamma \cF^j)  .}
Then one can show that 
\eqn\ComplicatedIdentityForMij {	\eqalign {
	& \left(1 + \Gamma \Delta \right)^{-1} 
	  \left( \frac 1 2 + \Gamma \cF^i \right) T_i 
		= (T_i^{-1} + T_j^{-1})^{-1}	\cr
	&= \left(1 + \Gamma \Delta \right)^{-1} 
	\left( \frac 1 2 - \Gamma \cF^j \right) T_j 
		= \left(M^{ij}\right)^{-1}, } }
where $M^{ij}$ has been defined in \eqr\TwoJMSymbol.
This allows us to find a simple expression for $X_c$ 
that we will use shortly:
\eqn\NormalFormForXij {
	M^{ij} X_c \equiv (E_1 +   E_2)  ,}
where
\eqn\DefinitionOfe	{
	E_1 = T^{-1}_i X_1 ,~~~ E_2 = T^{-1}_j X_2 .}
The $E$'s are convenient because 
\eqn\CanonicalCommutationFore { \eqalign { 
	\comm {E_2^\mu} {E_2^\nu} &= J^{\mu\nu} 
	= - \comm {E_1^\mu} {E_1^\nu}, \cr 
	\comm {E_1^\mu} {E_2^\nu} &= 0} }

	For $X_c$ to be defined, M has to be nondegenerate, which means 
$U_{ij}$ has no eigenvalue equal to $(-1)$.  Actually 
it might very well happen that for certain given $\cF_i$'s, 
for instance the $SU(2)$ case that we shall consider 
later, $U_{ij}$ does have eigenvalue to $(-1)$.  However, $U_{ij}$ 
is only defined up to $U_{ij} \to T_i S_i (T_j S_j)^{-1}$, 
where $S_i$ and $S_j$ are $Sp(D)$ transformations.    
It is easy to show that one can always find suitable $S$'s 
so that $M$ is nondegenerate.

\subsubsection {The product}

	In computing the actual product, the key step 
is the change of basis functions between the $aA$ and $aa$ 
representations.  We now 
find them for the $(i,j)$ string.   
We choose to diagonalize $E_1^a$ and $E_2^a$
in the $aa$ representation, $a$ ranging from $1$ to ${D\over 2}$.  
The 
first ${D\over 2}$ components of ${1\over 2}M^{ij} X_c$ is 
${1\over 2}\left(E_1^a + E_2^a \right)$. The canonical 
conjugates of the rest $D \over 2$ components 
are $\left(E_1^a - E_2^a \right)$.  
To find how the latter are represented, substituting  
the expressions for $\Delta$ and $X_c$ in 
\eqr\PIsCovariantDerivative, we find that 
\eqn\SolveForPi { \eqalign {
	\Pi &= P + {1\over 2}\left( \Delta + \cTheta \right) X_c \cr
	&= - \frac {M^\top J} 2  \left(E_1 - E_2 \right) \cr
	&  + \left( \cTheta 
	  + {T_i^{-1}}^\top J  T_j^{-1} 
	  - {T_j^{-1}}^\top J  T_i^{-1} 	\right)
	  	(2M)^{-1} \left(E_1 + E_2 \right) }}
Now we fix the gauge choice by requiring the second term in the 
last expression vanish.  Thus $(E_1 - E_2)^a$ are represented purely 
as derivatives with respect to 
$(M x)^A$.  Therefore the change of basis matrix element 
between the $(e_1^a, e_2^a)$ basis and the $X_c$ basis is 
\eqn\ChangeOfBasis {
	{\bra {e_1^a,e_2^a} \ket {x}} = 
	\frac  {\sqrt {\abs M}} {2^{D\over 2}} 
	\exp\paren{-\imath (e_1^a - e_2^a) J^{aA} ({1\over 2}M^{ij} x)^A}
	\delta \left(\frac {e_1^a + e_2^a} 2 
	- \frac {(M^{ij} x)^a} 2\right).}  
The determinant and powers of two appears as a result 
of  the different normalization between $X_c$ basis and the ${MX_c\over 2}$ 
basis.  They will not matter in the end.

	Now we are finally ready to compute the star product.  
\eqn\ComputationOfStarProduct { \eqalign {
	(\Psi^i_j * \Phi^j_k) (x) \propto &
	\myI {de^a_1 de^a_2}~ \bra {x} \ket {e^a_1, e^a_2} 
	(\Psi^i_j * \Phi^j_k) (e^a_1, e^a_2) \cr
	= & \myI {dx' dx''} \myI {de_1^a de_2^a dm_1^a dm_2^a} 
	 \Psi^i_j (x') \Phi^j_k (x'')  \delta (m_1^a - m_2^a) \cr
	& \bra {x} \ket {e^a_1, e^a_2} \bra {e^a_1, m_2^a} \ket {x'} 
\bra {m_1^a, e^a_2} \ket {x''} \cr
	= & \frac  {\sqrt {\abs {M^{ij} M^{jk} M^{ik}}}} {2^{D\over 2}}
	\myI {dx' dx''} \Psi^i_j (x') \Phi^j_k (x'') \cr
	& \exp\paren{\frac \imath 2 (x^\top (\cF^{ik;ij} x' - \cF^{ik;jk} 
x'') 
	+ x'^\top \cF^{ij;jk} x'' )}\cr
        = & 2^{D\over 2} \sqrt {\frac {\abs {M^{ik}}} {\abs {M^{ij} M^{jk}}}}
          \exp\paren{ \frac \imath 2\ppa{x'} \Omega_{i j; j k}\ppa{x''}}
          \Psi^i_j(x')\Phi^j_k(x'')\Bigg{\vert}^{x'=S^{ik}_{ij}x}_{x''=S^{ik}_{jk}
          x} . }}

	Again we fix the normalization by requiring the recovery of 
the usual matrix product when all the $\Omega_i$'s vanish.  
Hence we would get
\eqn\answer{ 
 (\Psi^i_j * \Phi^j_k)(x) = 
\exp\paren{ \frac \imath 2\ppa{{x'}^{\mu}} 
	\Omega^{\mu\nu}_{i j; j k}\ppa{{x''}^{\nu}}}
          \Psi^i_j(x')\Phi^j_k(x'')\Bigg{\vert}^{x'=S^{ik}_{ij}x,x''=S^{ik}_{jk}
          x} . }
For plane waves, this translate to 
\eqn\StarProductInFourierBasis {
	\exp\paren{\imath k_1 x} *_{ijk} \exp\paren{\imath k_2 x} = 
		\exp\paren{- \frac \imath 2 k_1^\top \Omega_{i j; j k}k_2} 
	\exp\paren{\imath (k_1^\top S_{ij}^{ik} + k_2^\top S_{jk}^{ik}) x},}
which is the desired product.

	By using \eqr\RelationOfSOmegaSymbol, one can show that 
\eqn\Associativity	{	\eqalign	{
      &	(\exp\paren{\imath k_1 x} *_{ijk} \exp\paren{\imath k_2 x}) 
		*_{ikl} \exp\paren{\imath k_3 x}
	= \exp\paren{\imath k_1 x} *_{ijl} (\exp\paren{\imath k_2 x}
		*_{jkl} \exp\paren{\imath k_3 x}),	\cr	
	&= \exp\paren{-{1\over 2} \imath ( k_1^\top \Omega_{i j; j k} k_2 
		+ k_2^\top \Omega_{j k; k l} k_3 
		+ k_1^\top \Omega_{i j; k l} k_3) }	
	\exp\paren{\imath (k_1^\top S_{ij}^{il} + k_2^\top S_{jk}^{il}
		+ k_3^\top S_{kl}^{il}) x}  }	}
thus proving associativity.

\section {The Case of $SU(2)$}

	In this section we deal with the simplest instance of
non-Abelian geometry: $N = 2$ and the deformation 
parameter is in the adjoint of $SU(2)$.  That is to say the 
noncommutativity parameter on 
brane $1$ is $\Omega$ 
but that on brane $2$ 
is $-\Omega$.  For this situation one may certainly apply 
the method developed in the previous section again.  
But we shall take this opportunity to consider a variation 
and illustrate the meaning of the large degree of freedom 
in choosing the geometric observables mentioned earlier.

	For simplification of notation, we can, by means 
of a congruence transformation, 
bring $\Omega$ to its canonical form $J$ and shall work in this 
basis till near the end of this section.  We call the coordinate 
observables on the left and right ends of the string 
$L^\mu$ and $R^\mu$ respectively.  Unlike the previous 
section, where $E_2$ is generically in a different parameterization 
of $R^D$ from $E_1$, related by some linear transformation, 
here $R$ is the same parameterization as $L$.  
Therefore $\comm {L^\mu} {L^\nu}$ and $\comm {R^\mu} {R^\nu}$ 
are exactly opposite in sign  
on a $11$ string, but identical 
on a $12$ string.

\subsection {Split Ordering}

	The Moyal-Weyl product can serve
as a method of quantization, i.e. mapping a function on 
the phase space (in our case, $R^D$) to an operator to 
the Hilbert space of a quantum mechanical model.  
As usual there is the ambiguity of operator ordering, 
and Moyal-Weyl product makes a symmetric choice.  
There are other orderings, and they can also be obtained 
by variation of the method developed in the \refs\ZYIN 
~ and reviewed in the last section.  Recall that to 
represent states in the LSQM 
Hilbert space as (wave) functions on $R^D$, we had to 
choose a set of $D$ geometric observables, simultaneously 
diagonalized in the $aA$ representation.  The action 
on them by the generator of translation  
should be what one expects for coordinates being 
translated.  In the last section, $X_c$'s are the geometric observables, 
but there are many other choices.  Some of them, giving 
different values to $\cTheta$, correspond to different choices of phase for 
the wave function.  Some other choices  corresponds to different 
operator ordering schemes in the language of quantization.  
Both will show up here.

	Let us divides the coordinates of the present 
problem into two 
groups which are canonical conjugates to each other with 
respect to $J$ (and $-J$).  We label them with $a, b, \ldots$ and 
$A, B, \ldots$ respectively as in section 2.1
Then we choose as geometric observables for any $(i,j)$ string 
\eqn\DefinitionOfMixedCoordinates { \eqalign {
	E^a &= L^a, \cr
	E^A &= R^A. }}

	Now let us consider the $11$ string.  First 
we will describe a scheme for illustration only 
that will not be used again in the paper.  Therefore 
to avoid confusion we use $\doteq$ instead of $=$ 
in equations peculiar to this example.
The basic commutation relations are 
\eqn\CanonicalCommuatationForPlusPlus { \eqalign {
	- \comm {L^\mu} {L^\nu} &= J^{\mu\nu} 
	= \comm {R^\mu} {R^\nu} , \cr
	\comm {L^\mu} {R^\nu} &= 0. } }
The translation operator is 
\eqn\TranslationOperatorForPlusPlus {
	P = - J(L - R). }
By a specific choice of phase of the basis state 
in the $aA$ representation, we can implement translation
by differentiation with respect to the space coordinates 
as per tradition: $P= \Pi$.  This means in particular that 
\eqn\UnusedRelationOfeRaWithem {
	R^a \doteq - \imath J^{aA} 
		\frac \pa {\pa{e^A}} + e^a. }	
Then by another choice of phase 
the change of basis between $aA$ and $aa$ basis 
is described by 
\eqn\UnusedChangeOfBasisBetweenaaAndaA {
	\bra {e} \ket {L^a, R^a} 
	\doteq \delta (e^a - L^a) 
	\exp^{\imath (R^a - L^a) J^{aA} e^A}. }
Then one finds that the noncommutative product is given 
by 
\eqn\UnusedPlusPlusProductWithPlusPlus {
	\Psi^1_1* \Phi^1_1 (e) 
	\doteq  e^{\imath \frac \pa {\pa M^A} 
		J^{Aa} \frac \pa {\pa M^a} } 
		\Psi^1_1(e^a, M^A) \Phi^1_1(M^a, e^A) 
		\Bigg{\vert}^{M^A = e^A}_{M^a = e^a}.  }
This corresponds, in quantization, to a choice of ordering 
in which all the $E^a$'s are brought to the 
left and all the $E^A$'s are brought to the right.

	However, in the rest of the paper we shall only 
use a variant of this ordering so that 
the condition $\alpha_1 = \alpha_2$ is satisfied 
in \eqr\DefinitionOfeA and the final result could 
be in a more convenient form.   
Another choice of phase in the $aA$ representation is 
made which replaces
\eqr\UnusedRelationOfeRaWithem by 
\eqn\RelationOfeRaWithem {
	R^a = - \imath J^{aA} 
		\frac \pa {\pa{e^A}}. }		
Then \eqr\UnusedChangeOfBasisBetweenaaAndaA is replaced by 
\eqn\ChangeOfBasisBetweenaaAndaA {
	\bra {e} \ket {L^a, R^a} 
	= \delta (e^a - L^a) 
	\exp\paren{\imath R^a J^{aA} e^A}, }
and \eqr\UnusedPlusPlusProductWithPlusPlus by 
\eqn\PlusPlusProductWithPlusPlus {
	\Psi^1_1* \Phi^1_1 (e) 
	=  \exp\paren{\imath \frac \pa {\pa M^A} 
		J^{Aa} \frac \pa {\pa M^a} } 
		\Psi^1_1(e^a, M^A) \Phi^1_1(M^a, e^A) 
		\Bigg{\vert}^{M^A = 0 = M^a}.  }
The $U(1)$ phase that relates this and the last one 
is given by the unit element in this new product.  
Instead of $1$, it is $\exp\paren{\imath e^a J ^{a A} e^A}$.  
We call this scheme {\em split ordering}.

\subsection {Off diagonal elements}

	On a 12 string the commutation relations are 
\eqn\CanonicalCommuatationForPlusPlus { \eqalign {
	\comm {L^\mu} {L^\nu} &= - J^{\mu\nu} 
	= \comm {R^\mu} {R^\nu} , \cr
	\comm {L^\mu} {R^\nu} &= 0. } }
The translation operator is 
\eqn\TranslationOperatorForPlusPlus {
	P = - J(L + R). }
A crucially new feature is that $P$ no longer commutes among 
themselves: $\comm {P^\mu} {P^\nu} = - 2 \imath J^{\mu\nu}.$
By a specific choice of gauge we can implement it
as 
\eqn\EquationForPInOneTwo {
	P_\mu= - \imath \frac \pa {\pa{e^\mu}} - [J e]_\mu.}
This means in particular 
\eqn\RelationOfeRaWithem {
	R^a = - \imath J^{aA} 
		\frac \pa {\pa{e^A}}. }	
Then by a choice of phase 
consistent with the split ordering 
the change of basis between $aA$ and $aa$ basis 
is described by 
\eqn\ChangeOfBasisBetweenaaAndaA {
	\bra {e^a, e^A} \ket {L^a, R^a} 
	= \delta (e^a - L^a) 
	\exp\paren{\imath R^a J^{aA} e^A}. }
Then one finds that the noncommutative product is given by 
\eqn\PlusPlusProductWithPlusPlus {
	\Psi^1_1 * \Phi^1_2 (e) 
	=  \exp\paren{\imath \frac \pa {\pa M^A} 
		J^{Aa} \frac \pa {\pa M^a} } 
		\Psi^1_1(e^a, M^A) \Phi^1_2(M^a, e^A) 
		\Bigg{\vert}^{M^A = 0 = M^a}.  }

	Using this method systematically we find all the 
possible products $\Psi^i_j *_{ijk} \Phi^j_k$.  They in fact 
can be written in a very compact matrix form:
\eqn\SUTWOProduct	{
	\paren {\Psi \times \Phi} (e^a, e^A)
	\equiv  \Psi(e^a, M^A) 
	\exp\paren{\imath \ppa {M^A} {\hat{\bf{\Omega}}}^{Aa} \ppa {M^a}} 
		\Phi(M^a, e^A) \Bigg{\vert}^{M^A = 0 = M^a},	}
where ${\hat{\bf{\Omega}}}^{Aa}  = J^{Aa} \sigma_3$ and 
all products are understood as matrix products\foot
	{$\sigma_3$ is just the usual element of Pauli matrices:
	\eqn\DefinitionOfsigmaThree {
		\sigma_3 = \left( \matrix {
		1 &	 0 \cr
		0 & 	-1  \cr} \right). }}.
This is highly suggestive of an $SU(2)$ valued Poisson 
structure ${\hat{\bf{\Omega}}}$.  For each $(\mu,\nu)$ pair, 
${\hat{\bf{\Omega}}}^{\mu\nu}$ is a two by two matrix, 
intuitively in the adjoint of $SU(2)$.  In this case, 
\eqn\ExpressionForOmega	{
	{\hat{\bf{\Omega}}}^{\mu\nu} = J^{\mu\nu} \sigma_3.}
This product is clearly associative.

	The unit element of this new product is 
\eqn\UnitElementForSUTwo	{
	\id_{su(2)}	= \exp\paren{- \imath e^a {\hat{\bf{\cF}}}_{aA}
 e^A}, }
where 
\eqn\DefinitionOfFHat {
	\hat{\bf{\cF}}_{\mu\nu} = - J^{\mu\nu} \sigma_3.}

\subsection {Non-Abelian Geometry}

	Just as in the general case discussed in the last section, 
constant matrices form a subalgebra.  
However, 
\eqn\CommutatorWithM	{
	\Psi \times M = \Psi(e^a, 0) M 
	\neq M \Psi(0, e^A) = M \times \Psi,}
unless $\Psi(e^a, 0) = \Psi(0, e^A)$ is a constant matrix 
that commutes with $M$.  Therefore one cannot obtain the whole algebra 
by tensoring this matrix subalgebra with some other algebra.
Curiously, there are two other {\em distinct} matrix subalgebras 
with the interesting properties:
\eqn\AnotherTwoMatrixSubalgebra		{	\eqalign {
	(M \suid) \times \Psi = M \Psi, \spsep
	\Psi \times (\suid M) = \Psi M, }}
so that the total algebra is a left and right module 
under them separately and respectively.

	The new algebra defined by \eqr\SUTWOProduct 
and \eqr\answer also contains 
subalgebras that are the deformations of that of the scalar 
functions on $R^D$.  However, there are $N$, rather than 
just one, of them, distinguished by their 
deformation parameters $\Omega_i$.  No one is more preferred 
than the others.  On the other hand, a well defined deformed 
algebra of functions is essential for the noncommutative 
geometric interpretation of D-brane worldvolume.  A noncommutative 
space is itself defined only by the algebra of functions ``on it.''
The loss of a canonical noncommutative algebra of scalar 
functions calls for 
a drastic reinterpretation of the underlying ``space.''   In the present 
case, the $N$ different algebras represent $N$ deformed 
noncommutative spaces on top of each other, distinguished only by 
their deformation parameters.  However, this simplistic picture overlooks 
all the $(i,j)$ strings.  Indeed, it is clearly not ``covariant'' 
enough.  
The total algebra is not a simple tensor product of any one of the 
scalar subalgebras with some matrix algebra, as it were for the usual 
case of Abelian or no deformation.  What is the meaning 
of this?

	We propose that these algebra define examples of 
a new type of geometry, which we call {\em non-Abelian Geometry}.  It is
a type quite apart from both the original underlying commutative space
$R^D$ and the noncommutative $R^D$ defined by the Moyal product because
the matrix (and more generally, non-Abelian) degree of freedom and the
function degree of freedom become entangled everywhere and become one
entity.  Recall that
the algebra of functions of the direct product of two manifolds, 
$M \times M'$, is the tensor product of their respective algebras: 
\eqn\TensorProductAlgebra {
	A_{M\times M'} = A_M \otimes A_{M'}.}
So should be the case for the 
direct product of noncommutative spaces.  Now also recall that 
the algebra of $N\times N$ matrices can be reinterpreted as the 
algebra of functions on a certain discrete noncommutative space: 
a discrete ``fuzzy'' torus with $N$ units of magnetic flux\foot 
	{This can be constructed, for example, on a two-torus as 
	follows.  With $N$ (more generally, rational) units of flux 
	through the torus, the algebra of functions contains centers.  
	One finds that the algebra has an $N$ 
	dimensional irreducible representation in which the 
	Fourier components are realized as clock and shift 
	operators of a $N$-ary clock and the products thereof.  The latter 
	generate the algebra of $N\times N$ matrices.}.
Therefore the usual case of $N\times N$ matrix fields on commutative or 
$U(1)$ deformed space can be reinterpreted as (the algebra on) 
the direct product of the continuous space with the appropriate discrete 
torus.  When the $U(N)$ bundle is nontrivial, it should be identified as 
a fibration of the discrete torus over the continuous base space.  
The notion of a base space is based on the existence of 
a canonical algebra of scalar valued functions on it, of which 
the total algebra is a module.  This semi-decoupling makes it a matter 
of taste whether to consider the total algebra as representative of 
a combined noncommutative space or just conventionally as an adjoint 
module of the algebra of the base space.  
However, with non-Abelian deformation considered in this 
paper, such a canonical scalar algebra ceases to exist.
The discrete torus and the continuous space therefore lose 
their independent identities and separate meaning.  
The conventional picture has to be replaced by a total space 
that intertwines the discrete, non-Abelian degrees of freedom of the Yang-
Mills theory
with that 
of the  continuous, noncommutative space.  
This is the precise meaning of {\em non-Abelian geometry}.

\section {The Taub-NUT connection} 

So far we have studied an example of Lorentz non-invariant theory. These
theories give new deformations to the otherwise constrained
structure of quantum field theory. As discussed above, they can be
realized in string theory when we have a 
background B-field. In the presence of branes we have basically four choices
of orienting the B-field resulting in four different theories. The first
case would be to orient the B-field transverse to the brane \refs\CDGR
i.e the B-field is polarized orthogonally. Naively such a constant B-field
can be gauged away. However if we also have a nontrivial orthogonal space 
$-$ say a Taub-NUT $-$ and one leg of the B-field is along the Taub-NUT 
cycle then this configuration 
give rise to new theories known as the pinned brane theories\refs\CDGR. 
The D-branes
have minimal tension at the origin of the Taub-NUT and therefore
the hypermultiplets in these theories are massive. The mass is given by
\eqn\masshyper{
m^2={b^2\over 1+b^2}}
where $b$ is the expectation value of the B-field at infinity. The origin
of the mass of the hypermultiplets is easy to see from the T-dual version. For 
simplicity we will take a D3 brane oriented along $x^{0,1,2,3}$ 
and is orthogonal to a Taub-NUT which has a non trivial metric 
along $x^{6,7,8,9}$. The coordinate
$x^6$ is the Taub-NUT cycle and the B-field
has polarization $B_{56}$. Making
a T-duality along the compact direction of the Taub-NUT we have a 
configuration of a NS5 brane and a D4 brane. The hypermultiplets
in this model come from strings on the D4 brane crossing the NS5 brane. 
Due to the twist on the torus $x^{5,6}$, the D4 on the NS5 brane comes back
to itself by a shift resulting in the hypermultiplets being massive while the 
vectors remain
massless. 

The second case is to orient the B-field with one leg along the brane
and the other leg orthogonal to it \refs\DGR. Again 
we could gauge away such a 
B-field. But in the presence of Taub-NUT $-$ with the leg of the B-field
along the Taub-NUT cycle $-$ we generate new theories on the brane 
known as the dipole theories\refs\BerGan.
 Hypermultiplets in these theories have dipole 
length
$L$ determined by the expectation of the B-field. The vector multiplets
have zero dipole lengths. The dipoles are light and typically the branes are
not pinned. The field theory on the branes are nonlocal theories with
the following  multiplication rule:
\eqn\multirule{
(\Phi\circ \Psi)(x) \equiv \exp\paren{{1\over 2}(L_1^i{\del\over \del x''^i}
-L_2^j{\del\over \del x'^j})}
\Phi(x')\Psi(x'')\Bigg{\vert}^{x'=x''=x}}
where $\Phi(x)$ and $\Psi(x)$ have dipole lengths $L_1$ and $L_2$
respectively. It is easy to check that when we specify the dipole 
length of the above product as $L_1+L_2$, the multiplication rule is
associative. The dipoles in these theories are actually rotating
arched strings stabilized (at weak coupling) by a generalized magnetic
force\refs\DGR. 
In this limit the radiation damping and the coulomb attraction
are negligible. Again the T-dual model can illustrate why the hypermultiplets 
have
dipole length. We take the above configuration of a D3 transverse to a 
Taub-NUT but now with a B-field $B_{16}$.
Under  T-duality we get a configuration of a NS5 brane and a
D4 brane with a twisted $x^{1,6}$ torus. Along direction $x^1$ the
D4 comes back to itself up to a twist. Since both the 
NS5 and the D4 branes are 
{\it along} $x^1$, this shift gives a dipole length to the hypermultiplets.
Observe that this way the vectors have zero dipole length. 

The third  case is to orient the B-field completely along the branes. Here
we cannot gauge away the B-field. Gauging will give rise to $F$ field on
the world volume of the branes. This would also
mean that now we no longer need any nontrivial
manifold. The supersymmetry will thus be maximal (in the above two cases the 
supersymmetry was reduced by half or less). 
The theory on the brane is noncommutative 
gauge
theory. 
The B-field modifies the boundary
conditions of the open strings describing the D branes.
This modification is crucial in giving a non zero
correlation function for three (and more) gauge propagators. This in turn
tells us that the usual kinetic term of the gauge theory is replaced by\foot{
 $g_{ij}$ is the open string metric $G_{ij}$ of \refs\SWNCG.}
\eqn\kinterm{
\int~\sqrt{det g}~g^{ii'}g^{jj'}F_{ij}\ast F_{i'j'}}
where \eqr\kinterm 
involves an infinite sequence of terms due to the definition of
$\ast$ product \eqr\StarProduct.
The above equation is written in terms of {\it local} variables i.e the 
variables defining the usual commutative Yang-Mills theory. The map which 
enables us to do this is found in
\refs\SWNCG. 
In other words, noncommutative YM at low energies can be viewed as a 
simple tensor deformation of commutative YM. This deformation is also
responsible in producing a scale $\Omega$ in the theory. A key 
feature is that 
this scale governs the size of
smallest lump of energy that can be stored in space. Any lump of size 
smaller than this will have more energy and therefore will not be physically 
stable. For a D3 brane with a B-field $B_{23}$ T-duality along $x^3$ will 
give a D2 brane on a twisted $x^{23}$ torus. The non locality in this picture
can be seen from the string which goes around the $x^3$ circle and reaches 
the D2 with a shift \refs\DH.  

For the dipole theories (which are again non-local theories) there also exist
a map with which we could write these theories in terms of local
variables\refs\BerGan. 
This map is relatively simpler than the Seiberg-Witten map for
noncommutative theories. Using this map one can show that the dipole
theories at low energies are simple vector deformations of SYM theory
\refs\DGR.

The fourth case is the topic of this paper. Here, as discussed earlier, we 
have a configuration of multiple D-branes with different $B$-fields 
oriented parallel to the branes. However one difference now is that this
configuration may or may not preserve any supersymmetry. Also the 
multiplication rule in this theory is more complex and now there is no
clear distinction between the non-Abelian and the noncommutative spaces. The
algebra \eqr\answer reflects this intertwining.  
	
{}From the above considerations it would seem that all the four theories 
have distinct origins. However, as we shall discuss below, all these theories
can be derived from a particular setup in M-theory but with different
limits of the background parameters. This will give us a unified way to
understand many of the properties of these theories. 

Consider first the pinned brane case. If we lift a D4 brane with 
transversely polarized B-field we will have a configuration of a M5 brane
near a Taub-NUT singularity and a C-field having one leg parallel to
the M5 brane. The limits of the external parameters which give rise to 
decoupled theory on the M5 brane are\refs\CDGR:
\eqn\limit{
C\to \epsilon,~~~R\to \epsilon,~~~M_p\to \epsilon^{-\beta},~~~\beta> 1}
where $R$ is the Taub-NUT radius and $M_p$ is the Planck mass. In this
limit the energy scale of the excitations of the M5 brane is kept finite
whereas the other scales in the problem are set to infinity.

The dipole theories are now easy to get from the above configuration. 
Keeping the background limits same we rotate the M5 brane such that the 
C-field now has two legs along the M5 (its still orthogonal to the Taub-NUT).
With this choice a simple calculation will tell us that that the M5 is 
not pinned in this case.
 
To generate the noncommutative gauge theories we first remove the M5 brane 
from the picture and identify the M-theory direction with the Taub-NUT 
circle. Now the limits\foot{These limits however don't specify the 
complete decoupling of the theory. This is because the theory has negative
specific heat\refs{\MALRUS}.} 
which give rise to $6+1$ dimensional
noncommutative gauge theories are\refs\DGRunp:
\eqn\ncglimit{
C\to \epsilon^{-1/2},~~~R\to fixed,
~~~M_p\to \epsilon^{-1/6},~~~g^M_{\mu\nu}\to \epsilon^{2/3}}
where $g^M_{\mu\nu}$ is the dimensionless M-theory metric\foot{There is an
interesting digression to the above cases. Between the two limits of the 
background $C$ field there exist a case under which 
$$C\to finite,~~R\to finite,~~M_p\to \infty,~~g_s\to 0$$ This gives us
another decoupled theory on the M5 brane which is in the same spirit as
the little string theory\refs\CDGR.}. 
This limit is 
the same as
Seiberg-Witten limit and the coupling constant of the theory
\eqn\coupcons{
g^2_{YM}= M_p^{-3}C = fixed.}

The theory of non-Abelian geometry 
can be  studied in M-theory  using a multi Taub-NUT 
background with a $G$-flux that has non-zero expectation values near the
Taub-NUT singularities. As discussed in \refs\DGS, such a choice of 
background flux generally breaks
 supersymmetry. From type IIA D6 brane point of 
view this flux 
will
appear as gauge fluxes $F_i = dA_i$ on the $i^{th}$
world volume\foot{There is a subtlety here. 
For F/M-theory compactification with G-flux, when there is a generic 
flux$-$ not concentrated near the singularities of the manifold $-$ this 
appears in the corresponding type IIB theory as $H_{NSNS}$ and $H_{RR}$
background. However when the flux is concentrated near the singularity, then
it appears as gauge fields on the brane\refs\DGS.}. 
 If the Taub-NUT is
oriented along $x^{7,8,9,10}$, $x^7$ being the Taub-NUT circle, and the $C$
field has two legs along the Taub-NUT and one leg along $x^1$ then the 
world volume gauge fields $A_i$ can be determined by decomposing the $C$ 
field as:
\eqn\aifromC{
C(t,y,x^7,{\vec r}) = \sum_{i=1}^N ~ A_i(t,y) \wedge {\cal L}^{(2)}_i(x^7,
{\vec r})}
where $y$'s are the coordinates $x^{1,..,6}$ of the $D6$ brane world volume,
${\cal L}^{
(2)}_i$ are the harmonic forms of the multi Taub-NUT and 
$\vert {\vec r} \vert = \sqrt {x^kx^k},~k=8,9,10$. 
However it turns out 
that these harmonic forms are {\it deformed} from their original values 
due to the background $G$-flux. Therefore in the above equation the
harmonic forms are $C$-twisted ones which preserve their anti-self-duality
with respect to the $C$-twisted background metric. The precise form of this
twist will be presented elsewhere \refs{\KRSav}.  
In this frame-work it 
might be possible $-$ by pure geometric means $-$ to see this 
intertwining more clearly.

Before we end this section let us summarize the connections
between different theories governed by the $B$ field dynamics in the following
table\foot{We have chosen to make everything non-Abelian. Its an 
straightforward exercise to extend the above three Abelian cases to this.}:

\ \medskip
\begintable
{\it Theories}|{\it SUSY}|{\it Product rule}:~$(\Phi^i_j*\Psi^j_k)(x)$\eltt
Pinned branes|$\cN=2$|$\Phi^i_j(x) \Psi^j_k(x)$\elt
Dipole theory|$\cN \le 2$|$e^{{1\over 2}(L_1^i{\del\over \del x''^i}
-L_2^j{\del\over \del x'^j})}~\Phi^i_j(x')\Psi^j_k(x'')\vert^{x'=x''=x}$\elt
Noncomm. geometry|$\cN=4$|$e^{\paren {\frac \imath 2
		 \frac \pa {\pa{x'^{\mu}}}\Omega^{\mu\nu}
		 \frac \pa {\pa{x''^{\nu}} }}}~
	\Phi^i_j(x') \Psi^j_k(x'') \vert^{x'' = x' = x}$\elt
Non-Abelian geometry|$\cN \le 1$|
$e^{\paren {\frac \imath 2
		\frac \pa {\pa{x'^{\mu}}}\Omega_{ij;jk}^{\mu\nu} 
		 \frac \pa {\pa{x''^{\nu}} }}}~
          \Phi^i_j(x')\Psi^j_k(x'')\vert^{x'=S^{ik}_{ij}x}_{x''=S^{ik}_{jk}
          x}$
\endtable

\section{Discussions and Conclusion}

In this section we will discuss possible supergravity background for the 
analysis presented in the earlier sections. We will also illustrate 
some aspects of this using a mode expansion for a background $U(1)\times
U(1)$. This case is related to the recent analysis done in \refs\DolNap, 
where the worldsheet propagator was calculated to compute two distinct 
noncommutativity parameter. It was shown that near one of the brane, say $1$,
the $\ast$-product involves $\Omega_1$ only. This is clear from our analysis
because $\Psi^1_1 * \Phi^1_1$ involves $\Omega_{11;11}$ which from 
\RelationOfSOmegaSymbol\ is just $\Omega_1$. 

Another recent paper which dealt with some related aspects is \refs\RT. Here
two different $\ast$ products arise naturally in the fractional brane setting.
In this model we have a configuration of $D5-\bar{D5}$ wrapping a vanishing
two cycle of a Calabi-Yau.
This model however is supersymmetric and for some special choice of
background $B$ and $F$ fields the tachyon is massless\refs\RT.

\subsection {Gravity Solutions}

Let us first consider the case of a large number of D3 branes on top of
each other and with a background B-field switched on. The B-field is constant
along the world volume of the D3 branes. What is the supergravity solution
for the system? Obviously the near horizon geometry cannot be AdS as there
is a scale $\Omega$ in the theory which breaks the conformal invariance.
Indeed, as shown by Hashimoto and Itzhaki\refs\HASI and independently by
 Maldacena and Russo\refs\MALRUS, the supergravity solution can be calculated
by making a simple T-duality of the D3 brane solution. Under a T-duality the 
B background becomes metric and it tilts the torus which the D2 brane wraps.
This solution is known and therefore we could calculate the metric for this 
case and T-dualize to get our required solution. Observe that under a 
T-duality we do get a B-field which is constant {\it along} the brane but is
a nontrivial function along the directions orthogonal to the brane. In other 
words there is a $H$ field. But for all practical purposes this solution 
is good enough to give us the near horizon geometry of the system. The 
scale of the theory appears in the metric deforming our AdS background which
one would expect in the absence of B-field. For the case in which 
D3 branes are along $x^0,x^1,.., x^3$ and B-field has a polarization $B_{23}$
the near horizon geometry looks like\refs{\HASI,\MALRUS}:
\eqn\nearhoriz{
ds^2 = \alpha'{\Big [}u^2(-dx_0^2+dx_1^2)+u^2h(dx_2^2+dx_3^2)+{du^2\over u^2}
+d\Omega_5^2{\Big ]}}
where $h=(1+a^4u^4)^{-1}$ and $a^2$ is the typical scale in the theory (it is
related to $\theta$).

The above metric has the expected behavior that for small $u$ the theory 
reduces to $AdS_5\times S^5$. From gauge theory this is the IR regime of the
theory. One naturally expects that noncommutative YM reduces to ordinary YM
at large distances. The above solution has an added advantage that
it tells us that below the scale $a$ (which will be proportional to
$\sqrt{\Omega}$) commutative variables are no longer the right parameters 
to describe the system accurately. Noncommutativity becomes the inherent
property of the system and therefore local variables fail to capture all the
dynamics.

At this point we should ask  whether our new system has a consistent 
large N behavior. The gauge theory is highly noncommutative of course 
but it also has a large number of noncommutativity parameters (typically N).
We have a system of N D3 branes with gauge fields $F^i, i = 1,..N$ on them. 
We can simplify the problem by taking only one polarization of the gauge 
fields, i.e we would concentrate on the fields $F_{23}^i$. 

A simple analysis tells us immediately that the previous procedure to 
generate a solution is not going to help in this case. The procedure is
suitable to generate {\it one} scale and therefore we should now rely on 
different technique.
 Also the system now has no supersymmetry and therefore
we have to carefully interpret the background. 

Let us denote the magnetic field $F^i_{23}$ on the branes as ${\cal B}_i$. 
We can make a Lorentz transformation to generate a {\it constant} magnetic
field $\cal B$ on the branes but different electric fields ${\cal E}_i$
such that the relations $${\cal B}_i^2 = {\cal B}^2-{\cal E}_i^2, ~~~
{\cal E}_i\bullet B=0$$ are 
satisfied.
We can also make a gauge transformation to convert the constant magnetic 
field to a background constant B-field. 

We now make a T-duality along the $x^3$ direction\foot{There could be a 
subtlety in performing a 
T-duality here because the string theory background is not
supersymmetric. But we are considering a T-duality completely from the
supergravity point of view in which the transformation of the bosonic
background is important for us. As such the extra corrections are 
not relevant for studying this.}. Under this the electric
fields ${\cal E}_i$ will become velocities $v_i$ of the D2 branes and the 
B-field will tilt the torus $x^2,x^3$ as before. therefore the final
configuration will be a bunch of D2 branes (or, in a reduced sense, points)
moving with velocities $v_i$ along the circle $x^3$. 

At this point it would seem that the supergravity solution is easy to 
write down. But there are some  subtleties here.
Recall that when we had a single scale $\Omega$ in the problem and 
the T-dual picture was a D2 brane wrapped on a tilted torus, T-duality along
$x^3$ was easy because we had assumed that the harmonic function of the
D2 brane is {\it delocalized} along the third direction. Therefore the D2 
brane is actually {\it smeared} along that direction. This typically has 
the effect that the harmonic function of the D2 brane is no longer 
$1+{Q_2\over r^5}$ rather its $1+{Q_2\over r^4}$, $Q_2$ is the charge of the
D2 brane. This is the case that we have to consider. Delocalizing the D2 
branes would mean that we have an infinite array of D2 branes moving with 
velocity, say, $v_1$ and so on. 
Also since the system lacks 
supersymmetry the velocities are {\it not} constant. An interpretation 
of this model can be given from fluid mechanics. Due to delocalization
we have layers of fluid moving with velocities $v_i$ along $x^3$ with a 
viscosity between them. This would tend to retard the motion of the various 
layers making the problem slightly nontrivial.
But as we shall see some interesting property of 
the system is obvious without going to the original (T-dual) model.

The metric for the D2 brane (for simplicity the direction 23 are on a 
square torus) is given by 
\eqn\dtwo{
ds^2= H^{-1/2}ds^2_{012}+H^{1/2}ds^2_{34..9}}
where the harmonic function satisfy:

$$ \del^2 H = \Sigma_{i=1}^N~~\delta(r_i)$$
and $r_i$ is given by $r_i=\sqrt{(x_3-v_it)^2+x_4^2+...+x_9^2}$ when the 
velocities are small so that we could neglect relativistic effects. 
Recall that the system is delocalized along direction $x_3$ therefore there
are actually an infinite array of branes moving(i.e it behaves like a fluid). 

Let $r=\sqrt{x_4^2+...+x_9^2}$ then its easy to show that for a large radius
of $x_3$ circle and near horizon geometry (i.e. $r\to 0$) the harmonic 
function is modified from the naive expected value. The harmonic function
becomes:
\eqn\harmonica{
H(r)= 1+\sum_{i=1}^N~{1\over r^4}\int_0^{\pi/2}~{\sin^3\theta\over 
	(1-{v_it\over r} \sin\theta)^5}~d\theta}
when the compact direction is very small one can show that we get 
$H(r)= 1+{1\over r^4}$ . 
A T-duality along that direction will 
give us a noncompact D3 brane whose harmonic function will have the right 
property. This calculation is done without assuming any force between the 
branes. A more detailed analysis would require the behavior of open
strings between the branes. In the next  section we will elaborate on 
this issue by doing a mode expansion.

\subsection{Mode Expansion}

For simplicity we will take two D3 branes having fluxes $F_i=F^{(i)}_{23}$ 
on them. The D3 branes are oriented along $x^{0,1,2,3}$ and let 
$z=x^2+ix^3$ such that the mode expansion for the system becomes:
\eqn\modeexp{
z=\sum_n~A_{n+\nu}\exp\paren{(n+\nu)t}\Big[\exp\paren{i(n+\nu)\sigma}+
\Big({1-iF_1\over 1+iF_1}\Big)\exp\paren{-i(n+\nu)\sigma}\Big]}
The quantity $\nu$ measures the shift in the mode number due to the presence
of different gauge fluxes at the boundary. This shift can be easily worked 
out from the boundary conditions at the two ends of the open string. In 
terms of the above variables $\nu$ is given by
\eqn\nuvalue{
\nu={1\over 2\pi}~\sin^{-1}\Big[2{(F_2-F_1)(1+F_1F_2)\over (1+F_1^2)
(1+F_2^2)}\Big] }
As is obvious from the above formula when the gauge fluxes are same on the
different branes we do not expect any shift in the mode number. This shift
can now be used to calculate the new ground state energy of the system. This
zero-point energy, in the NS sector, will now depend on $\nu$. Using the
identity
\eqn\identity{
\sum_{n\ge 0}~(n+\nu)= -{1\over 12}(6\nu^2-6\nu+1)}
the zero-point energy can be calculated from the bosons, fermions and the
ghosts contributions. The bosons and the fermions along directions $x^{2,3}$
are quantized  with mode numbers $n+\nu$ and $n\pm 
\vert\nu-{1\over 2}\vert$ 
respectively. In general for a system of $Dp$ branes with fluxes $F_{1,2}$
the zero point energy is given by
\eqn\ladoo{
-{1\over 2\alpha'}\Big({p-1\over 4}+\Big{\vert}
 \nu-{1\over 2}\Big{\vert}\Big)}
{}From the above formula its clear that there is a tachyon on D3 brane for
any values of $\nu$. For very small values of $\nu$ the tachyon has 
$m^2=-{1\over 2\alpha'}(1-\nu)$ and for large values of $\nu$ it has 
$m^2=-{\nu\over 2\alpha'}$. Also now there is a subtlety about GSO projection.
Therefore it depends whether we study a $D1$ brane or $\bar{D1}$
brane. When the branes are kept far apart then there would be no tachyon
in the system but the branes will be attracted to each other which in turn
will retard the velocities of the brane.

Let us now consider the special case of $SU(2)$. For this we have the 
following background
\eqn\sutwo{
F_1=-F_2=-F}
It is straightforward to show that now the modes will be shifted by $\nu$
given as
\eqn\modeshift{
\nu={2\over \pi}~tan^{-1}F}
For the case we are interested in, $F\to \infty$, and therefore the shift
$\nu=1$. The ground state energy do not change but all the modes of the string
get shifted by 1. 

It would be worthwhile to analyze in greater detail the spectrum 
and dynamics of this theory.

\vskip.5in

\noindent{\bf Acknowledgments:} 
\bigskip
K.D. would like to thank J.~Maldacena, S.~Mukhi and R.~Tatar for useful 
discussions.
Z.Y. would like to thank D.~Gaitsgory and C.~H.~Yan for useful 
discussions.  We would like to thank N.~Seiberg for helpful comments.
The work of K.D. is supported by DOE grant number
DE-FG02-90ER40542 and the work of Z.Y. is supported by NSF grant number 
PHY-0070928.

\listrefs    
\end